\documentclass[twocolumn, superscriptaddress, preprintnumbers, amsmath, amssymb, aps]{revtex4}

\allowdisplaybreaks
\allowdisplaybreaks[4]
\usepackage{CJK}
\usepackage{graphicx}
\usepackage{dcolumn}
\usepackage{bm}
\usepackage{anyfontsize}
\usepackage{mathptmx}
\DeclareMathAlphabet{\mathcal}{OMS}{cmsy}{m}{n}

\usepackage[
            pdfstartview=FitH,
            CJKbookmarks=true,
            bookmarksnumbered=true,
            bookmarksopen=true,
            colorlinks,
            linkcolor=blue,
            anchorcolor=blue,
            citecolor=blue
            ]{hyperref}

\begin{document}

\title{Entanglement in two-quasiparticle-triaxial-rotor systems:
Chirality, wobbling, and the Pauli effect}

\author{Q. B. Chen} 
\email[Contact author:~]{qbchen@phy.ecnu.edu.cn}
\affiliation{Department of Physics, East China Normal University,
Shanghai 200241, China}

\author{S. Frauendorf}
\email[Contact author:~]{sfrauend@nd.edu}
\affiliation{Physics Department, University of Notre Dame, Notre
Dame, IN 46556, USA}

\date{\today}

\begin{abstract}

We investigate the entanglement in two-quasiparticle plus triaxial-rotor (PTR) 
model for the particle-hole configuration $\pi(1h_{11/2})^1 
\otimes \nu(1h_{11/2})^{-1}$, the particle-particle configuration 
$\pi(1h_{11/2})^1 \otimes \nu(1h_{11/2})^1$, and 
two-proton particles configuration  $\pi(1h_{11/2})^2$ for different
values of the triaxiality parameter. The entanglement between 
the angular momenta of the two quasiparticles and 
the total angular momentum is quantified by 
the three bipartite concurrences $\mathcal{C}$ of one type 
of angular momentum with the other two angular momenta and   
the area $\mathcal{F}$ of the triangle formed by the bipartite 
concurrences. Collective chiral and wobbling modes 
are identified for $\gamma>15^\circ$ via spin 
coherent state (SCS) maps and spin squeezed state (SSS) plots. 
Their entanglement increases from moderate values at the band 
head to near-maximal values at $I=20$. The area $\mathcal{F}$ 
of the chiral partners changes order as function of $I$ which 
reflects the crossing of the partner bands  as a signature of 
chirality. For the $\pi(1h_{11/2})^2$ configuration, the 
antisymmetrization required by the Pauli exclusion principle
causes strong entanglement between the two 
protons, which significantly amplifies the area $\mathcal{F}$. 
For $\gamma<15^\circ$, the lowest bands become various uniformly 
rotating quasiparticle configurations, which have large 
values of $\mathcal{F}$ for all values $I$. 
 
\end{abstract}

\maketitle

\section{Introduction}
\label{intro}

Nuclear chirality arises in rapidly rotating nuclei 
with a triaxially deformed core, where high-$j$ valence 
particle(s) and hole(s) are crucial~\cite{Frauendorf1997NPA}. 
In the body-fixed frame, particle(s) align along the short ($s$) axis, 
hole(s) along the long ($l$) axis, and core along the 
medium ($m$) axis. This arrangement can break chiral symmetry, 
leading to degenerate left- and right-handed states. 
In the laboratory frame, quantum tunneling restores 
symmetry by exchanging angular momentum, resulting in 
nearly degenerate chiral doublet bands~\cite{Frauendorf1997NPA} 
with $\Delta I = 1$ and identical parity, 
first observed in four $N = 75$ isotones~\cite{Starosta2001PRL}.
Furthermore, the multiple chiral doublet (M$\chi$D) 
bands within individual nuclei with different 
configuration~\cite{J.Meng2006PRC} or identical 
configuration~\cite{Droste2009EPJA, Q.B.Chen2010PRC, 
Hamamoto2013PRC} represent a significant extension of 
the chiral symmetry concept, illustrating its multifaceted 
manifestations in nuclear structure.

The experimental observation of over 50 chiral doublet 
bands or M$\chi$D in mass regions around $A \approx 80$, 100, 
130, and 190 highlights the widespread occurrence of this 
phenomenon in nuclear physics. Extensive reviews~\cite{J.Meng2010JPG, 
J.Meng2014IJMPE, Bark2014IJMPE, J.Meng2016PS, Raduta2016PPNP, 
Starosta2017PS, Frauendorf2018PS, Q.B.Chen2020NPN, 
S.Y.Wang2023FoP, Budaca2024FoP, Bark2024FoP, Jolos2024FoP, 
Grodner2024FoP, Petrache2024book} provide a thorough 
overview of this research field, including detailed 
data tables~\cite{B.W.Xiong2019ADNDT}. Correspondingly, 
various theoretical approaches have been developed
to investigate the chiral doublet bands. For example, the particle plus
triaxial-rotor model (PTR)~\cite{Frauendorf1997NPA, Starosta2002PRC, Koike2003PRC, 
J.Peng2003PRC, Koike2004PRL, S.Q.Zhang2007PRC, B.Qi2009PLB, Lawrie2010PLB, 
Q.B.Chen2018PRC, Q.B.Chen2018PRC_v1, Q.B.Chen2018PLB, 
Q.B.Chen2019PRC, Y.Y.Wang2019PLB, Y.Y.Wang2020PRC, Q.B.Chen2020PLB} 
and its approximate solution~\cite{Raduta2014JPG, 
Raduta2016JPG, Raduta2017JPG, Budaca2018PRC_v1, Budaca2019PLB, Budaca2021PLB}, 
the titled axis cranking (TAC) model~\cite{Dimitrov2000PRL, 
Olbratowski2004PRL, Olbratowski2006PRC, P.W.Zhao2017PLB, Y.P.Wang2023PLB}, the 
TAC plus random-phase approximation (RPA)~\cite{Almehed2011PRC}, the TAC 
plus the collective Hamiltonian method~\cite{Q.B.Chen2013PRC, Q.B.Chen2016PRC, 
X.H.Wu2018PRC}, the interacting boson-fermion-fermion model~\cite{Brant2008PRC}, 
the angular momentum projection (AMP) method~\cite{Bhat2012PLB, 
F.Q.Chen2017PRC, Shimada2018PRC_v1, Y.K.Wang2019PRC, Y.K.Wang2024PLB},
as well as the time-dependent relativistic density functional 
theory~\cite{Z.X.Ren2022PRC, B.Li2024PLB}. 

The aplanar chiral mode appears when the rotational frequency 
exceeds a critical value, $\hbar \omega_c$, as described in 
the three-dimensional TAC model~\cite{Olbratowski2004PRL, 
Olbratowski2006PRC, P.W.Zhao2017PLB, P.W.Zhao2019PRC, 
J.Peng2020PLB_v1, J.Peng2022PRC, D.Chen2023EPJA, 
Y.P.Wang2023PLB}. Below this critical frequency, angular momentum 
remains confined to the intrinsic $s$-$l$ plane (planar rotation), 
while above $\hbar \omega_c$, a transition to aplanar rotation 
occurs, manifesting chiral symmetry. In the PTR framework,
which treats the total angular momentum as a good quantum number, 
a critical spin $I_c$ marks the minimum spin needed for 
stable aplanar rotation. The mode evolves from chiral 
vibration (CV), characterized by the oscillation of the
total angular momentum with respect to the $s$-$l$ plane, to 
chiral rotation (CR), characterized by the aplanar rotation, 
with increasing angular momentum~\cite{Frauendorf1997NPA, Starosta2002PRC, 
B.Qi2009PRC, H.Zhang2016CPC, Q.B.Chen2019PRC, 
Budaca2019PLB, Q.B.Chen2020PLB, B.Hu2024PRC, Q.B.Chen2024PRC, 
Y.Wu2024PLB, Y.Wu2025PRC}. Observables, including $g$-factor~\cite{Grodner2018PRL,
Grodner2022PRC, Q.B.Chen2024PRC} and spectroscopic quadrupole 
moment~\cite{Q.B.Chen2020PLB, B.Hu2024PRC}, have validated 
the critical spin in systems like $^{128}$Cs~\cite{Grodner2018PRL, 
Grodner2022PRC}. 

In the PTR, chiral modes arise from the interaction between 
high-$j$ particles and holes, acting as gyroscopic degrees 
of freedom, and the triaxial rotor core. The CV to CR 
transition is driven by strong coupling between the angular 
momenta of the particle, hole, and core, leading to significant 
angular momentum entanglement. This makes the PTR ideal for 
studying entanglement and its implications in chiral phenomena, 
providing a clear understanding of quantum correlations
and angular momentum entanglement in a simple tripartite system. 

Entanglement is a fundamental concept in quantum mechanics 
that describes the non-factorizable correlations between 
subsystems of a composite quantum system, which cannot 
be fully characterized by the independent states of 
its components. In quantum many-body systems, entanglement 
manifests in specific signatures that are of significant 
interest in the fields of condensed matter physics and 
quantum field theory, where it provides insights 
into the structure and dynamics of complex 
systems~\cite{Calabrese2004JSM, Amico2008RMP, 
Peschel2009JPA, Horodecki2009RMP, Nishioka2009JPA, 
Eisert2010RMP, Lin2020NPB}. Recent advancements in quantum 
information theory and quantum computing have revitalized 
interest in the study of entanglement in nuclear 
physics~\cite{Enyo2015PRC, Enyo2015PTEP, Enyo2015PTEP_v1, 
Legeza2015PRC, Robin2021PRC, Faba2021PRA, 
Kruppa2022PRC, Pazy2023PRC, D.Bai2022PRC, Lacroix2022PRD, 
Tichai2023PLB, Bulgac2023PRC, Bulgac2023PRC_v1, Johnson2023JPG, 
C.Y.Gu2023PRC, Q.B.Chen2024PRC_v2, Jafarizadeh2024NPA,
B.Li2024PRC, Ghapanvari2025NPA, S.Y.Liang2025PRC, Brokemeier2025PRC,
Beane2019PRL, D.Bai2024PRC}, 
as it offers a powerful framework for understanding correlations
and quantum coherence in nuclear many-body systems. 

Several entanglement measures based on the density matrix 
are commonly employed to quantify many-body correlations 
in quantum many-body systems. One such measure is the von 
Neumann (vN) entropy, which quantifies the degree of quantum 
entanglement between two subsystems within a composite 
quantum system. The vN entropy has been extensively utilized 
in a wide range of studies concerning entanglement, particularly 
in condensed matter physics and quantum field 
theory~\cite{Calabrese2004JSM, Amico2008RMP, Peschel2009JPA, 
Horodecki2009RMP, Nishioka2009JPA, Eisert2010RMP, Lin2020NPB}. 
In addition, it has found significant applications in the 
study of atomic nuclei~\cite{Enyo2015PRC, Enyo2015PTEP, 
Enyo2015PTEP_v1, Legeza2015PRC, Robin2021PRC, 
Faba2021PRA, Kruppa2022PRC, Pazy2023PRC, D.Bai2022PRC,
Lacroix2022PRD, Tichai2023PLB, Bulgac2023PRC, Bulgac2023PRC_v1, 
Johnson2023JPG, C.Y.Gu2023PRC, Q.B.Chen2024PRC_v2, Jafarizadeh2024NPA,
B.Li2024PRC, Ghapanvari2025NPA, S.Y.Liang2025PRC}. For example, 
the authors of Ref.~\cite{Q.B.Chen2024PRC_v2} used vN entropy 
to study entanglement between valence quasiparticles 
and a triaxial rotor within the PTR model for $^{135}$Pr 
(one-quasiparticle) and $^{130}$Ba (two-quasiparticle), 
focusing on the coupling between total and quasiparticle 
angular momenta. By employing Schmidt decomposition, they 
quantified the entanglement via the entropy of 
subsystems, finding that entropy and entanglement
increase with spin $I$ and wobbling quanta $n$. However, 
the vN entropy is limited to bipartite systems and cannot 
describe the entanglement in the tripartite chiral mode.

Furthermore, the concurrence $\mathcal{C}$, introduced by Hill and 
Wootters~\cite{Hill1997PRL, Wootters1998PRL}, provides an important 
measure of entanglement for a general pair of qubits that is faithful: 
strictly positive for entangled states and vanishing for all separable 
states. In Ref.~\cite{Bhaskara2017QIP}, the concurrence was extended  
to multiparticle pure states in arbitrary dimensions.
For an arbitrary pure tripartite state $|\phi\rangle_{ABC}$
shared by three parties $A$, $B$, and $C$, the concurrence 
between the bipartition $A$ and $BC$ is 
\begin{align}
 \mathcal{C}_{A(BC)}=\sqrt{2[1-\textrm{Tr}(\rho_A^2)]},
\end{align}
with $\rho_A=\textrm{Tr}_{BC}(|\phi\rangle_{ABC}~_{ABC}\langle \phi|)$. 
It is an entanglement between part $A$ and the rest of the system $BC$, 
known as one-to-other bipartite entanglement. The state is biseparable 
as $A$ and $BC$ if and only if $\mathcal{C}_{A(BC)}=0$. The concurrence  
satisfies the following relationship 
\begin{align}
 \mathcal{C}_{A(BC)} \leq \mathcal{C}_{B(CA)}+\mathcal{C}_{C(AB)},
\end{align}
and its permutations with respect to the three parties $A$, $B$, and $C$. 
This relation suggests that entanglement owned by one party is no 
larger than the sum of entanglement by the other two.
Based on the concurrence, Ref.~\cite{S.B.Xie2021PRL} proposed a 
tripartite entanglement measure, which is related to 
the area of a so-called concurrence triangle, named as 
\textit{concurrence fill}. But, it was subsequently 
pointed that this measure is increasing under local operations and
classical communications (LOCC)~\cite{X.Z.Ge2023PRA}, 
which means it is not a proper entanglement measure.
Furthermore, a new proper genuine multipartite
entanglement measures are constructed by using the 
geometric mean area of these concurrence triangles~\cite{Z.X.Jin2023RP},
which are non-increasing under LOCC. Hence, we will 
employ the concurrence triangle area to study the entanglement 
in the chiral mode with particle-hole configuration. 

In this work, the entanglement of the chiral mode will be studied
for the particle-hole configuration $\pi(1h_{11/2}) \otimes 
\nu(1h_{11/2})^{-1}$. The effects of triaxial deformation 
on the entanglement will be discussed. Furthermore, for comparisons, 
we will investigate the entanglements in the systems of 
two-quasiparticle pairs $\pi (1h_{11/2})^1\otimes \nu(1h_{11/2})^1$ 
and $\pi (1h_{11/2})^2$ coupled with a triaxial rotor. 
In particular, the effects of the Pauli exclusion principle 
on entanglement will be studied. 

\section{Theoretical framework}

\subsection{Particle plus triaxial-rotor model}

In this work, the calculations are conducted within the 
framework of the PTR model. Considering a system comprising one proton 
and one neutron coupled to a triaxial collective 
rotor, the Hamiltonian within the PTR is expressed 
as follows~\cite{Bohr1975}
\begin{flalign}\label{eq:HPTR}
 \hat{H}_{\textrm{PTR}}=\hat{H}_{\textrm{coll}}+\hat{H}_p+\hat{H}_n.
\end{flalign}
Here, $\hat{H}_{\textrm{coll}}$ denotes the Hamiltonian of the  
rotor, expressed as:
\begin{align}\label{eq:Hcoll}
\hat{H}_{\textrm{coll}}=\sum_{k=1}^{3}\frac{\hat{R}^2_k}{2\mathcal{J}_k}
=\sum_{k=1}^{3}\frac{(\hat{J}_k-\hat{j}_{pk}-\hat{j}_{nk})^2}{2\mathcal{J}_k},
\end{align}
where the index $k=1$, 2, 3 corresponds to
the three principal axes of the body-fixed frame. The
$\hat{J}_k$, $\hat{R}_k$, $\hat{j}_{pk}$, and $\hat{j}_{nk}$ 
represent the angular momenta corresponding to the total nucleus, 
the collective rotor, the valence proton, and the valence neutron, 
respectively, and $\mathcal{J}_k$ are the 
three principal moments of inertia of the rotor.

Furthermore, $\hat{H}_{p(n)}$ represents the individual Hamiltonian 
of a  single proton (neutron) in 
the single-$j$ shell approximation \cite{Q.B.Chen2024PRC_v2}
\begin{align}\label{eq:eq2}
\hat{H}_{p(n)} =\frac{\kappa}{2}\Big\{\cos \gamma\Big[\hat{j}_3^2-\frac{j(j+1)}{3}\Big]
               +\frac{\sin \gamma}{2\sqrt{3}}\Big(\hat{j}_+^2+\hat{j}_-^2\Big)\Big\}.
\end{align}
Here, $\gamma$ denotes the  parameter measuring the triaxiality
of the mean field potential,
and the coupling parameter $\kappa$ is directly proportional to 
the quadrupole deformation parameter $\beta$ of it.

The Hamiltonian (\ref{eq:HPTR}) can be further 
decomposed as 
\begin{align}
 \hat{H}_{\textrm{PTR}}
  &=\hat{H}_{\textrm{rot}}
   +\hat{H}_{\textrm{rec}}^{(pp)}
   +\hat{H}_{\textrm{rec}}^{(nn)} 
   +\hat{H}_p+\hat{H}_n \notag \\
  &+\hat{H}_{\textrm{rec}}^{(pn)}
   +\hat{H}_{\textrm{cor}}^{(Ip)}  
   +\hat{H}_{\textrm{cor}}^{(In)},
\end{align}
with the rotational operator of the composed system
\begin{align}
 \hat{H}_{\textrm{rot}}=\sum_{k=1}^3 \frac{\hat{J}_k^2}{2{\cal J}_k},
\end{align}
the recoil terms 
\begin{align}
 \hat{H}_{\textrm{rec}}^{(pp)}
   &=\sum_{k=1}^3\frac{\hat{j}_{pk}^2}{2{\cal J}_k}, \quad
 \hat{H}_{\textrm{rec}}^{(nn)}
    =\sum_{k=1}^3\frac{\hat{j}_{nk}^2}{2{\cal J}_k}, \\
 \hat{H}_{\textrm{rec}}^{(pn)}
   &=\sum_{k=1}^3\frac{\hat{j}_{pk}\hat{j}_{nk}}{{\cal J}_k},  
\end{align}
and the Coriolis interaction terms
\begin{align}\label{eq:Cor}
 \hat{H}_{\textrm{cor}}^{(Ip)}
   =-\sum_{k=1}^3\frac{\hat{J}_k \hat{j}_{pk}}{{\cal J}_k}, \quad 
 \hat{H}_{\textrm{cor}}^{(In)}
   =-\sum_{k=1}^3\frac{\hat{J}_k \hat{j}_{nk}}{{\cal J}_k}.
\end{align}
Here, $\hat{H}_{\textrm{rot}}$ acts only on the orientation degrees of freedom
of the composite system, i.e., $\hat J$, while
$\hat{H}_{\textrm{rec}}^{(pp)}$ and $\hat{H}_{\textrm{rec}}^{(nn)}$ 
act only on the orientation of  the valence proton and neutron 
with respect to the body-fixed axes, i.e., $\hat{j}_{pk}$ and $\hat{j}_{nk}$.
Furthermore, $\hat{H}_{\textrm{rec}}^{(pn)}$ couples the degrees 
of freedom of valence proton and neutron, and 
$\hat{H}_{\textrm{cor}}^{(Ip)}$ and $\hat{H}_{\textrm{cor}}^{(In)}$ 
couples the rotational degrees of freedom of the total system  to 
the degrees of freedom of the valence proton and neutron, respectively. 
The entanglement of the rotation of the total system and the quasiparticles 
involves an entanglement between the rotor and quasiparticle degrees of freedom. 
To keep language simple, we will speak about ``rotor-quasiparticle" entanglement 
when showing the results for the entanglement  between the rotation
of the total composite system and the quasiparticles.

The entanglement among the rotor, valence proton, and 
valence neutron are generated by the $\hat{H}_{\textrm{rec}}^{(pn)}$,
$\hat{H}_{\textrm{cor}}^{(Ip)}$, and $\hat{H}_{\textrm{cor}}^{(In)}$. 
It is worth noting that $\hat{H}_{\textrm{rec}}^{(pn)}$
is repulsive, whereas $\hat{H}_{\textrm{cor}}^{(Ip)}$ 
and $\hat{H}_{\textrm{cor}}^{(In)}$ are attractive. 

Furthermore, to account for the pairing correlations, we employ
the standard BCS method in the PTR following Refs.~\cite{Ragnarsson1988HI,
S.Q.Zhang2007PRC}. In detail, we first diagonalize the single-$j$ shell
single-particle Hamiltonian (\ref{eq:eq2}) to obtain the single 
particle energies $e_v$ and single particle states $|v\rangle$ 
for the considered $j$-shell. According to the BCS method, the 
corresponding energies for the quasiparticle states are
\begin{align}
 \varepsilon_v=\sqrt{(e_v-\lambda)^2+\Delta^2},
\end{align}
and the occupation ($v_v$) and un-occupation ($u_v$) factors are
\begin{align}
 v_v &=\frac{1}{\sqrt{2}}\left[1-\frac{e_v-\lambda}{\sqrt{(e_v-\lambda)^2+\Delta^2}}\right]^{1/2}, \\
 u_v &=\frac{1}{\sqrt{2}}\left[1+\frac{e_v-\lambda}{\sqrt{(e_v-\lambda)^2+\Delta^2}}\right]^{1/2},
\end{align}
 where $\lambda$ denotes the Fermi energy and $\Delta$
the pairing gap parameter. To construct the matrix of the PTR Hamiltonian,
from the one excluding pairing, the single-particle energies
$e_v$ should be replaced by quasiparticle energies $\varepsilon_v$,
and each single-particle angular momentum matrix element
between $|v\rangle$ and $|v^\prime\rangle$ needs to be multiplied
by a pairing factor $u_{v^\prime} u_v +v_{v^\prime} v_v$~\cite{Ragnarsson1988HI,
S.Q.Zhang2007PRC}. 

\subsection{Reduced density matrix}

The PTR Hamiltonian (\ref{eq:HPTR}) can be solved through 
diagonalization within the strong-coupling basis. The Hilbert
space of the PTR model is the direct product of the Hilbert 
spaces of the three subsystems, $\mathcal{H}_{K\Omega_p\Omega_n}
=\mathcal{H}_K\otimes \mathcal{H}_{\Omega_p} \otimes \mathcal{H}_{\Omega_n}$.
The corresponding PTR eigenfunctions $|IM\rangle$ are written as 
\begin{align}\label{eq:eq3}
    |I M\rangle=\sum_{K\Omega_p\Omega_n} 
     f_{IK\Omega_p\Omega_n} |IMK\rangle \otimes |j_p\Omega_p\rangle \otimes 
     |j_n\Omega_n\rangle.
\end{align}
In these expressions, $I$ denotes the total angular momentum 
quantum number of the odd-odd nuclear system, inclusive of 
the rotor, proton, and neutron. The symbol $M$ signifies 
the projection onto the $z$ axis (3-axis) in the laboratory 
frame, while $K$ indicates the projection onto the intrinsic 
(body-fixed) frame's 3-axis. Furthermore, $\Omega_{p(n)}$ 
corresponds to the quantum number representing the 3-axis component 
of the valence nucleon angular momentum operator $j_{p(n)}$ 
in the intrinsic frame. The states $|IMK\rangle$ are 
represented  by the normalized Wigner 
functions $\sqrt{\frac{2I+1}{8\pi^2}}D^{I*}_{M,K}(\psi^\prime, 
\theta^\prime, \phi^\prime)$, which depend on three Euler angles. 
Adhering to the $\textrm{D}_2$ symmetry 
of a triaxial nucleus necessitates certain constraints on the 
values of $K$ and $\Omega_p$. Specifically, $K$ ranges from 
$-I$ to $I$, while $\Omega_p$ spans from $-j_p$ to $j_p$. 
As for $\Omega_n$, it varies between $-j_n$ and $j_n$. 
It is further required to satisfy the condition that 
$K_R = K-\Omega_p-\Omega_n$ is a positive even integer. 
In addition, one-half of all coefficients 
$f_{IK\Omega_p\Omega_n}$ is fixed by the symmetric relation
\begin{align}
 f_{I-K-\Omega_p-\Omega_n}=(-1)^{I-j_p-j_n} f_{IK\Omega_p\Omega_n}.
\end{align}
The coefficients $f_{IK\Omega_p\Omega_n}$ are determined by 
diagonalizing the Hamiltonian operator $\hat{H}_{\textrm{PTR}}$,
which provides the energy levels and associated wave 
functions of the system.

From the expanding coefficients $f_{IK\Omega_p\Omega_n}$ 
of the PTR wave functions (\ref{eq:eq3}), the matrix elements of 
the reduced density matrices for the total ($\rho^I$),
proton ($\rho^{j_p}$), and neutron ($\rho^{j_n}$) 
angular momenta can be constructed as
\begin{align}\label{eq:TWrhoj}
 (\rho^I)_{K K^\prime} &=\sum_{\Omega_p\Omega_n} f_{IK\Omega_p\Omega_n}^* 
 f_{IK^\prime \Omega_p\Omega_n},\\
 (\rho^{j_p})_{\Omega_p \Omega_p^\prime} &=\sum_{K\Omega_n} f_{IK\Omega_p\Omega_n}^* 
 f_{IK\Omega_p^\prime \Omega_n},\\
 (\rho^{j_n})_{\Omega_n \Omega_n^\prime} &=\sum_{K\Omega_p} f_{IK\Omega_p\Omega_n}^* 
 f_{IK\Omega_p \Omega_n^\prime}.
\end{align}
The reduced density matrices contain the information about
the distribution and correlations of angular momenta within
the system, which are the basis for interpreting the properties
and behavior of triaxial nuclei by means of the PTR.
In the following, we will study the entanglement of 
the system based on the reduced density matrices.

\subsection{Angular momentum geometry}

From the PTR wave functions one can study the underlying angular momentum 
geometry for the considered system to justify the existence of the chiral mode.

First of all, we calculate the probability distribution for 
the angular momentum orientation on the unit sphere projected 
on the polar angle $(\theta)$ and azimuthal angle $(\phi)$ 
plane, i.e., the spin coherent state (SCS) maps~\cite{Frauendorf2015Conf, 
Q.B.Chen2022EPJA} (also called as azimuthal plots in Refs.~\cite{F.Q.Chen2017PRC, 
Q.B.Chen2018PRC_v1, Streck2018PRC}). For the total angular 
momentum, it is calculated by the reduced density 
matrices $(\rho^I)_{K K^\prime}$ (\ref{eq:TWrhoj}) as~\cite{Q.B.Chen2022EPJA}
\begin{align}
 P(\theta, \phi) 
 &=\frac{2I+1}{4\pi}\sin\theta \notag\\
 &\times \sum\limits_{K,K^\prime=-I}^I D^{I*}_{IK}(0, \theta, \phi)
 (\rho^I)_{K K^\prime} D^{I}_{IK^\prime}(0, \theta, \phi).
 \label{eq:PSCS}
\end{align}
Obviously, the probability $P(\theta, \phi)$ satisfies 
the normalization condition
\begin{align}
 \int_0^\pi d\theta & \int_0^{2\pi} d \phi~P(\theta, \phi)=1.
 \label{eq:PSCSn}
\end{align}

Then, we calculate the probability distribution of 
the spin squeezed states (SSS) as given by Ref.~\cite{Q.B.Chen2024PRC_v1} 
\begin{align}\label{eq:PSSS}
 P(\phi)= \frac{1}{2 \pi}
  \sum\limits_{K,K^\prime=-I}^Ie^{-i(K-K^\prime)\phi}(\rho^I)_{K K^\prime},
\end{align}
where $\phi$ is the angle of the projection of the total angular 
momentum onto the short-medium ($s$-$m$) plane with the $s$ axis.
Clearly, the probability $P(\phi)$ also satisfies 
the normalization condition
\begin{align}
 \int_0^{2\pi} d \phi~P(\phi)=1.
 \label{eq:PSSSn}
\end{align}
In the case of axial symmetry with $\gamma=0^\circ$, 
$P(\phi)$ takes the constant value $1/2\pi$.

Furthermore, the mean square expectation values of the projections 
on the principal axes of the total, proton, and neutron angular momenta 
$\bm{J}$, $\bm{j}_p$, and $\bm{j}_n$ are calculated by~\cite{Q.B.Chen2022EPJA}
\begin{align}\label{eq:ANGPTR}
\langle \hat{J}^2_{k}\rangle &=\sum_{K K^\prime}
    (\rho^I)_{K K^\prime}\hat{J}^2_{k; K^\prime K}, \\
\langle \hat{j}^2_{pk}\rangle &=\sum_{\Omega_p \Omega_p^\prime}
    (\rho^{j_p})_{\Omega_p \Omega_p^\prime}\hat j^2_{pk; \Omega_p^\prime \Omega_p}, \\
\langle \hat{j}^2_{nk}\rangle &=\sum_{\Omega_n \Omega_n^\prime}
    (\rho^{j_n})_{\Omega_n \Omega_n^\prime}\hat j^2_{nk; \Omega_n^\prime \Omega_n}.
\end{align}
The $\hat{J}^2_{k; K^\prime K}$,
$\hat j^2_{pk; \Omega_p^\prime \Omega_p}$, and  
$\hat j^2_{nk; \Omega_n^\prime \Omega_n}$ are the corresponding
matrix elements of the total, proton, and neutron angular momenta, 
respectively. 

With the obtained angular momentum components $\langle \hat{J}_k^2\rangle$, 
$\langle \hat{j}_{pk}^2\rangle$, and $\langle \hat{j}_{nk}^2\rangle$,
one can calculate the expectation values of the corresponding 
orientation components along the $s$ axis 
$o_s=\langle \sin^2\theta\cos^2\phi\rangle$, the 
$m$ axis $o_m=\langle \sin^2\theta\sin^2\phi\rangle$, and the 
$l$ axis $o_l=\langle \cos^2\theta\rangle$ through the 
following relationships~\cite{Q.B.Chen2020PLB}
\begin{align}
\label{eq:OriI}
 o_k^I &=\frac{\langle \hat{J}_k^2\rangle +(I+1)/2}{(I+1)(I+3/2)}, \\
\label{eq:Orip}
 o_k^{p} &=\frac{\langle \hat{j}_{pk}^2\rangle +(j_p+1)/2}{(j_p+1)(j_p+3/2)}, \\
\label{eq:Orin}
 o_k^{n} &=\frac{\langle \hat{j}_{nk}^2\rangle +(j_n+1)/2}{(j_n+1)(j_n+3/2)}, 
\end{align}
for total, proton, and neutron angular momenta, respectively. 
Since the angular momenta are conserved in PTR, 
$o_s+o_l+o_m=1$ always holds. 

\subsection{Concurrence triangle area}

In this work, we use concurrence triangle area to characterize
the entanglement of two-quasiparticle coupled with triaxial rotor system. 
According to the definition in Ref.~\cite{Z.X.Jin2023RP}, 
the concurrence area for the present system can be calculated as 
\begin{align}
\label{eq:ConArea}
 \mathcal{F}_{Ij_pj_n}&=\Big[\frac{16}{3}
        \mathcal{Q}(\mathcal{Q}-\mathcal{C}_{I(j_p j_n)})
         (\mathcal{Q}-\mathcal{C}_{j_p(j_n I)})
         (\mathcal{Q}-\mathcal{C}_{j_n(Ij_p)})\Big]^{1/2},
\end{align}
with
\begin{align}
 \mathcal{Q}&=\frac{1}{2}\left(\mathcal{C}_{I(j_pj_n)}+\mathcal{C}_{j_p(j_nI)}
  +\mathcal{C}_{j_n(Ij_p)}\right), \\
\label{eq:ConI}
 \mathcal{C}_{I(j_pj_n)} &=\sqrt{2\Big\{1-\textrm{Tr}[(\rho^{I})^2]\Big\}}/
   \sqrt{2-\frac{2}{2I+1}},\\
\label{eq:Conjp}
 \mathcal{C}_{j_p(j_nI)} &=\sqrt{2\Big\{1-\textrm{Tr}[(\rho^{j_p})^2]\Big\}}/
   \sqrt{2-\frac{2}{2j_p+1}},\\
\label{eq:Conjn}
 \mathcal{C}_{j_n(Ij_p)} &=\sqrt{2\Big\{1-\textrm{Tr}[(\rho^{j_n})^2]\Big\}}/
   \sqrt{2-\frac{2}{2j_n+1}}.
\end{align}
Here, $\mathcal{Q}$ is the half-perimeter, and, thus, equivalent to the global
entanglement. The $\mathcal{C}_{I(j_pj_n)}$, $\mathcal{C}_{j_p(j_nI)}$, and 
$\mathcal{C}_{j_n(Ij_p)}$ are the three bipartite concurrences~\cite{Hill1997PRL, 
Wootters1998PRL, Bhaskara2017QIP} for $I(j_p j_n)$, $j_p(j_n I)$, 
and $j_n(Ij_p)$ partitions, respectively. The denominator in the 
expression of $\mathcal{C}$ and prefactor $16/3$ in $\mathcal{F}_{Ij_pj_n}$ 
are introduced to ensures the normalization $0\leq \mathcal{C}\leq 1$ and 
$0\leq \mathcal{F}_{Ij_pj_n}\leq 1$, respectively. The relevance of this
quantifier is that a state is biseparable if and only 
if $\mathcal{C}=0$. One notes that if $\hat{H}_{\textrm{cor}}^{(Ip)}
=\hat{H}_{\textrm{cor}}^{(In)}=0$, a $\bm{J}$ state can be separated from 
$\bm{j}_p$ and $\bm{j}_n$ states, so $\mathcal{C}_{I(j_pj_n)}=0$.
If $\hat{H}_{\textrm{cor}}^{(In)}=\hat{H}_{\textrm{rec}}^{(pn)}=0$,
$\mathcal{C}_{j_n(Ij_p)}=0.739$ due to the Kramer's degeneracy
for the neutron. Similarly, If $\hat{H}_{\textrm{cor}}^{(Ip)}
=\hat{H}_{\textrm{rec}}^{(pn)}=0$, $\mathcal{C}_{j_p(j_n I)}=0.739$.
Hence, the minimal $\mathcal{C}_{I(j_pj_n)}$ is zero, while 
the minimal $\mathcal{C}_{j_n(Ij_p)}$ and $\mathcal{C}_{j_p(j_n I)}$
are 0.739.

\section{Numerical details}

In the calculations, the quadrupole deformation parameters are taken as
$\beta \approx 0.25$ (corresponds to $\kappa=0.30~\textrm{MeV}/\hbar^2$) 
and $\gamma$ ranging from $30^\circ$ to $0^\circ$ in steps 
of $5^\circ$. To study a particle-hole configuration, the proton 
and neutron Fermi surfaces are taken as $\lambda_p=e_1$ and 
$\lambda_n=e_6$, respectively, where $e_1$, ...., $e_6$ are the egenstates 
of $\hat{H}_{p(n)}$ (\ref{eq:eq2}) in ascending order.  Namely, the
proton lying in the lower $h_{11/2}$ shell to plays the particle role
and neutron lying in the upper $h_{11/2}$ shell to plays the hole role.
The pairing gaps have the fixed values of $\Delta_p
=\Delta_n=1.0~\textrm{MeV}$. For the rotor part in PTR Hamiltonian, 
moments of inertia of the irrotational flow type, expressed 
as $\mathcal{J}_k=\mathcal{J}_0 \sin^2(\gamma-2k\pi/3)$
with $\mathcal{J}_0=30~\hbar^2/\textrm{MeV}$, are utilized. 

\section{Results and discussions}

\subsection{Energies}

\begin{figure}[ht]
  \begin{center}
    \includegraphics[width=0.90\linewidth]{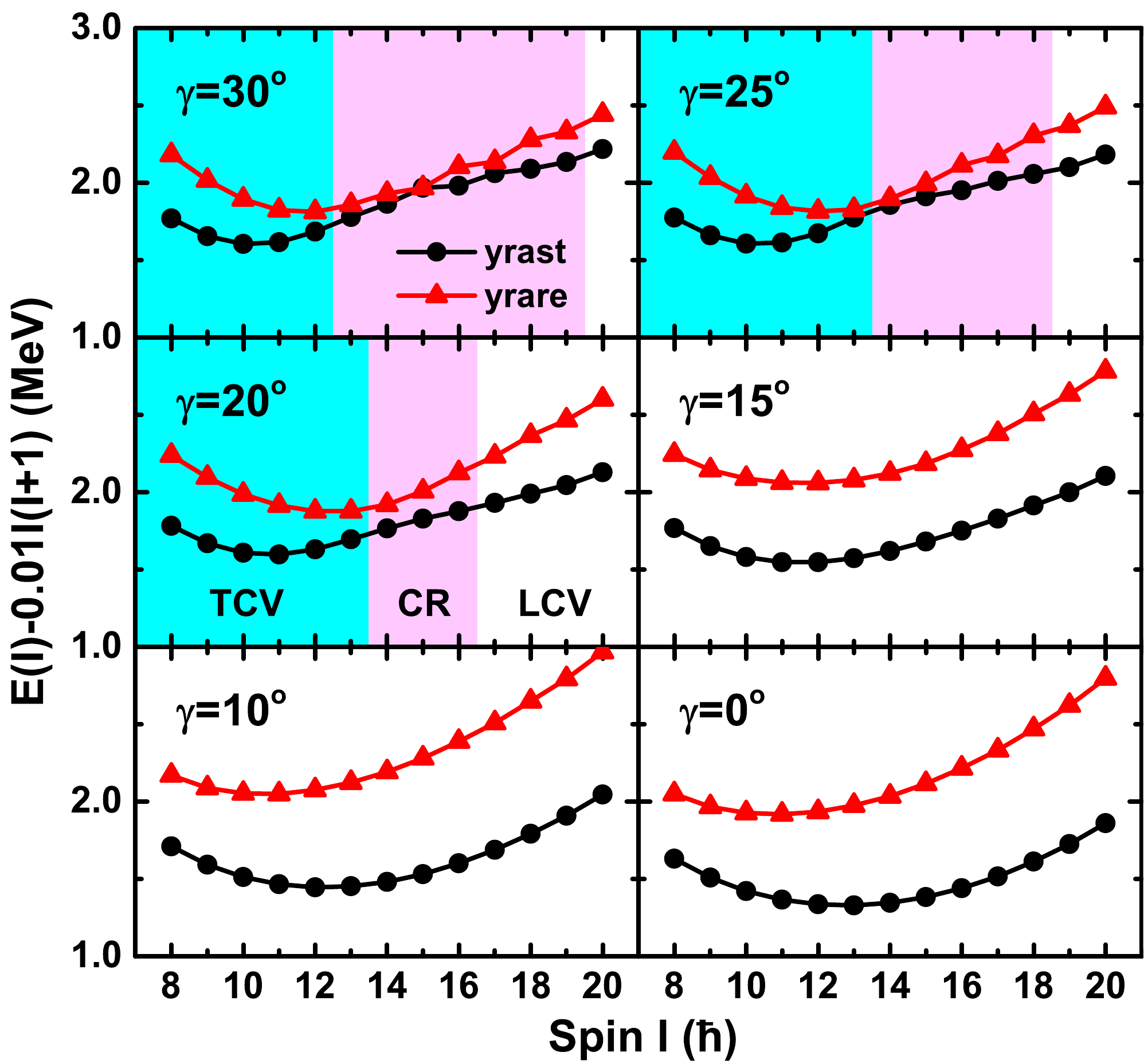}
    \caption{(Color online) Energies $E(I)$ 
    of the yrare and yrast bands as functions of spin $I$
    with $\gamma$ ranging from $30^\circ$ to $0^\circ$ 
    in step of $5^\circ$. The shadows denote the 
    regions of transverse chiral vibration (TCV) and chiral 
    rotation (CR), respectively.}\label{f:Energy}
  \end{center}
\end{figure}

\begin{figure}[t]
  \begin{center}
    \includegraphics[width=0.90\linewidth]{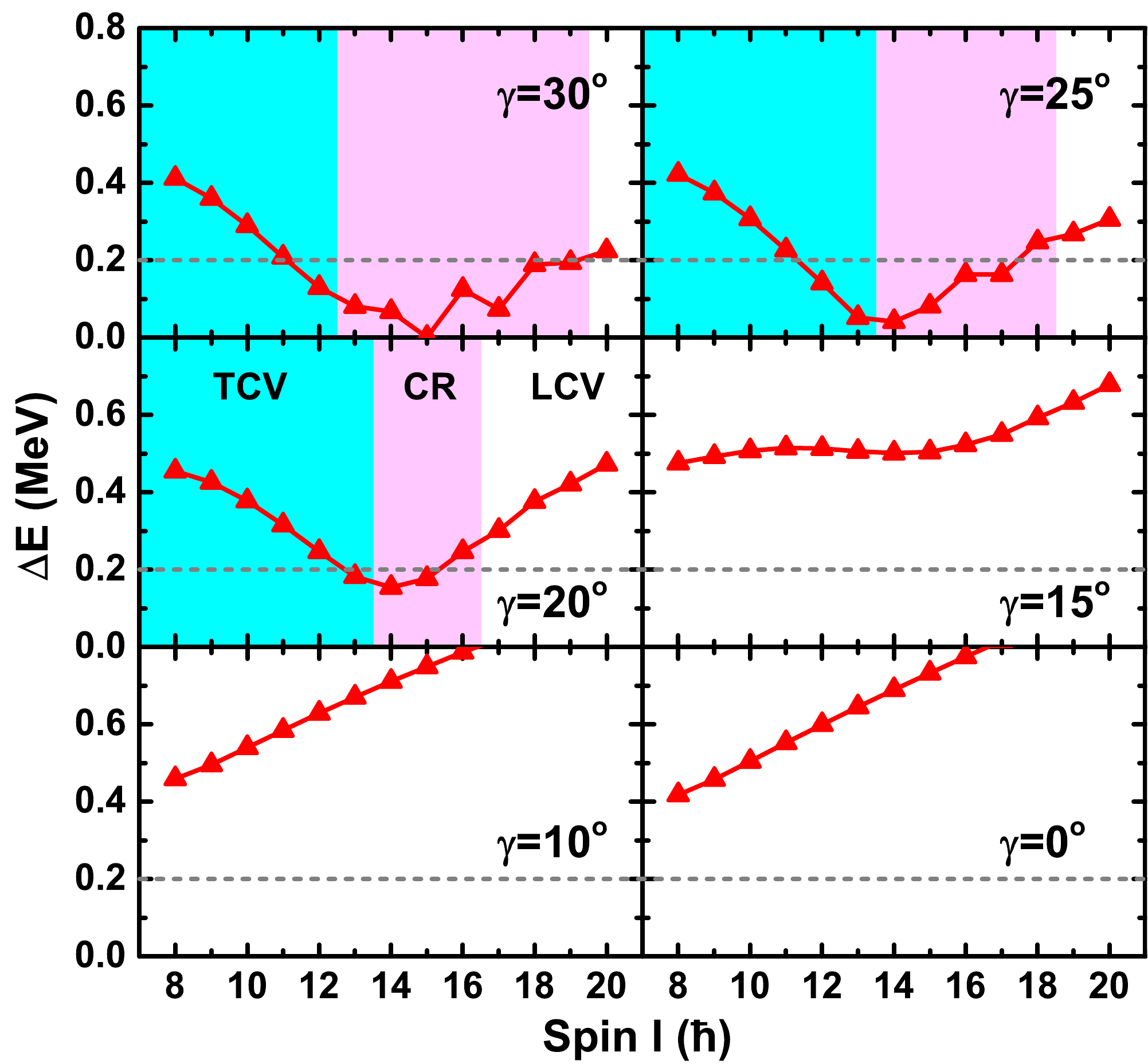}
    \caption{(Color online) Energy splittings $\Delta E(I)$ 
    between the yrare and yrast bands as functions of spin $I$
    with $\gamma$ ranging from $30^\circ$ to $0^\circ$ 
    in step of $5^\circ$. The dashed line in each panel labels 
    the position of $\Delta E(I)=0.2~\textrm{MeV}$. The shadows 
    denote the regions of transverse chiral vibration (TCV) 
    and chiral rotation (CR), respectively.}\label{f:DeltaE}
  \end{center}
\end{figure}

Figure~\ref{f:Energy} shows the energies of the yrast 
and yrare states. Figure~\ref{f:DeltaE} displays the energy 
splitting $\Delta E(I)=E_{\textrm{yrare}}(I)-E_{\textrm{yrast}}(I)$ 
between two states, which is used to assess the degree 
of chiral symmetry breaking. The smaller $\Delta E(I)$ 
the stronger the chiral symmetry is broken. The horizontal 
lines $\Delta E=0.2~\textrm{MeV}$ serve delineating 
the regions of ``static chirality" (chiral rotation) as follows. 
The average time to flip from the left-handed configuration 
to the right handed is $h/2\Delta E$. The rotational 
frequency around $I=15$ is about $\hbar \omega=0.4~\textrm{MeV}$.
That is, the nucleus stays for two turns in the left-handed 
configuration before flipping to the right-handed configuration.  
The figures also indicate the regions chiral vibration and 
chiral rotation by color shadows, where the criterions are
based on the angular geometry as explained below. 

The calculated $\Delta E(I)$ as a function of spin $I$ with 
$\gamma$ ranging from $30^\circ$ to $0^\circ$ in step
of $5^\circ$ are depicted in Fig.~\ref{f:DeltaE}. 
For $\gamma=30^\circ$, $25^\circ$, and $20^\circ$, as 
the spin increases, the inter-band energy splitting
$\Delta E(I)$ gradually decreases towards a value close 
to 0 MeV, before increasing again. The smallest 
$\Delta E(I)$ occurs at $I=15$, $14$, and 
$14$ for $\gamma=30^\circ$, $25^\circ$, and $20^\circ$,
respectively. For $\gamma = 15^\circ$, $\Delta E(I)$ starts at a 
relatively high value at the lower spin states and
remains constant for $I \leq 15$, after which 
it increases. The large value of $\Delta E(I)$
suggests that chiral doublet bands do not form for this 
value of $\gamma$ and smaller values. Comparing 
$\gamma = 10^\circ$ with $\gamma = 0^\circ$, $\Delta E(I)$ suggests
that the triaxial deformation has a minor effect on the system. 

\begin{figure*}[!ht]
  \begin{center}
    \includegraphics[width=0.95\linewidth]{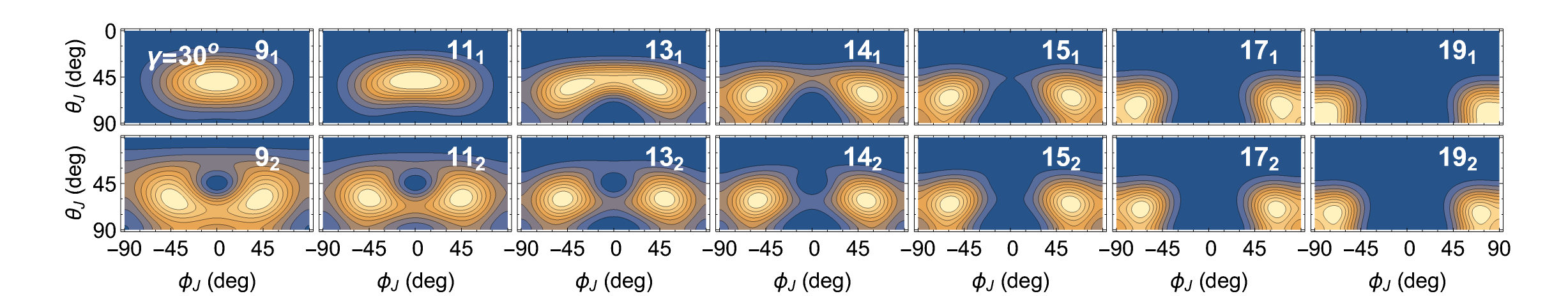}\\
    \includegraphics[width=0.95\linewidth]{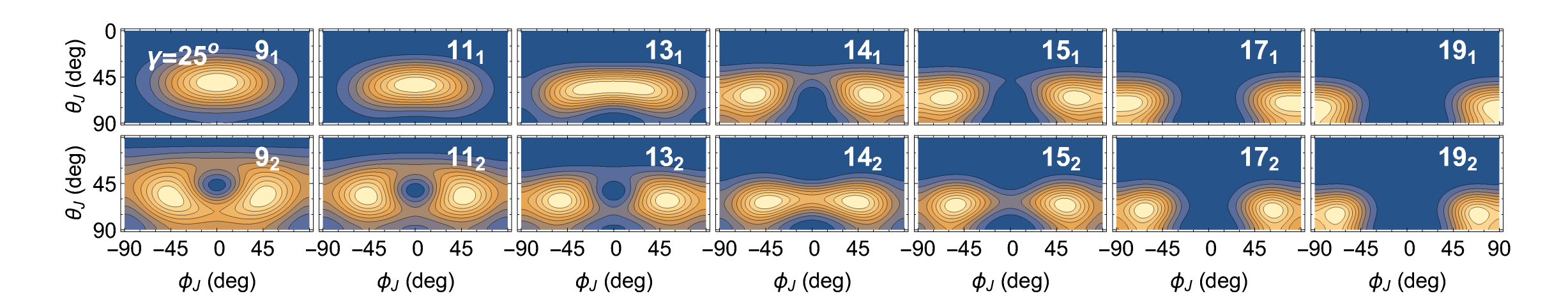}\\
    \includegraphics[width=0.95\linewidth]{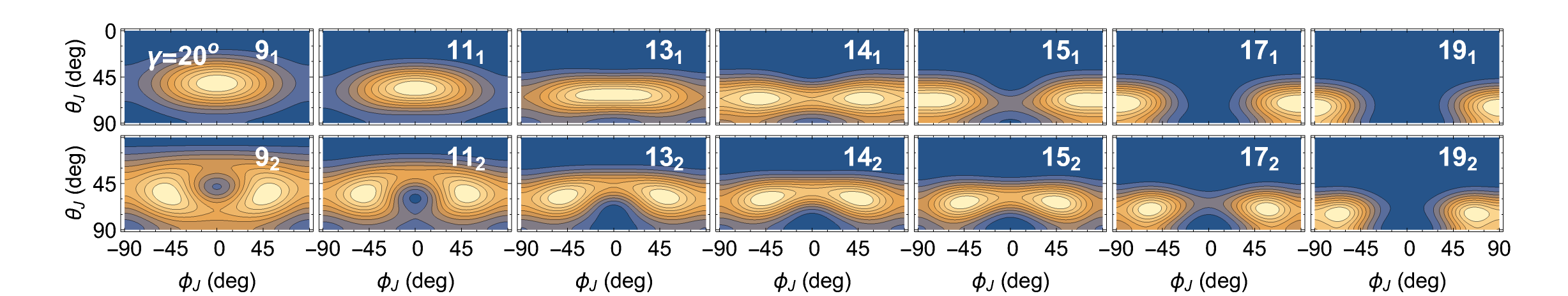}\\
    \includegraphics[width=0.95\linewidth]{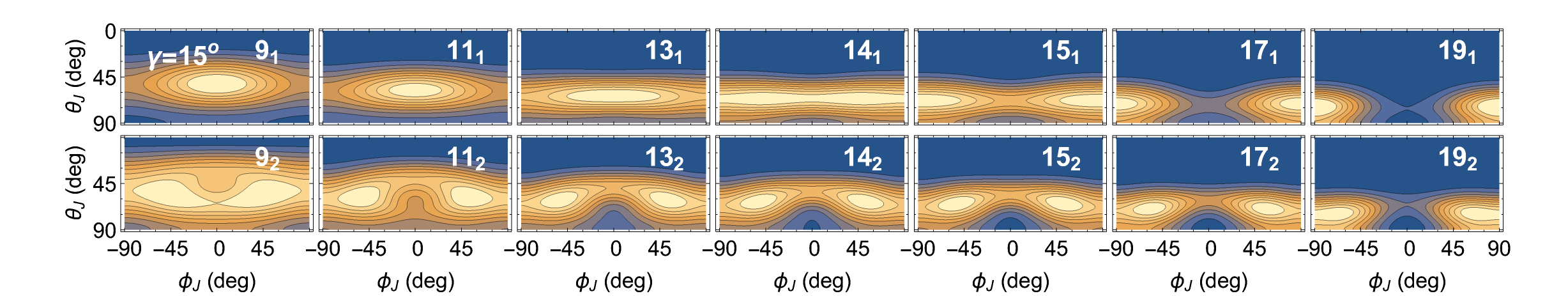}\\
    \includegraphics[width=0.95\linewidth]{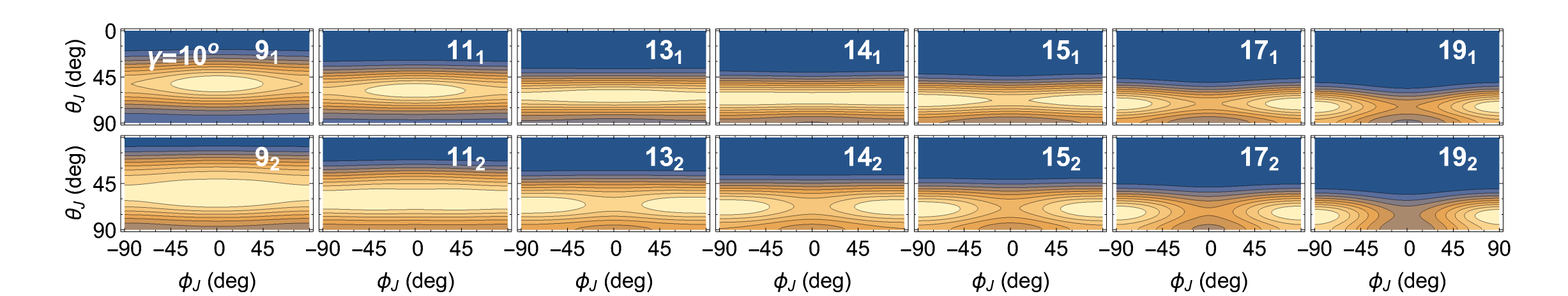}\\
    \includegraphics[width=0.95\linewidth]{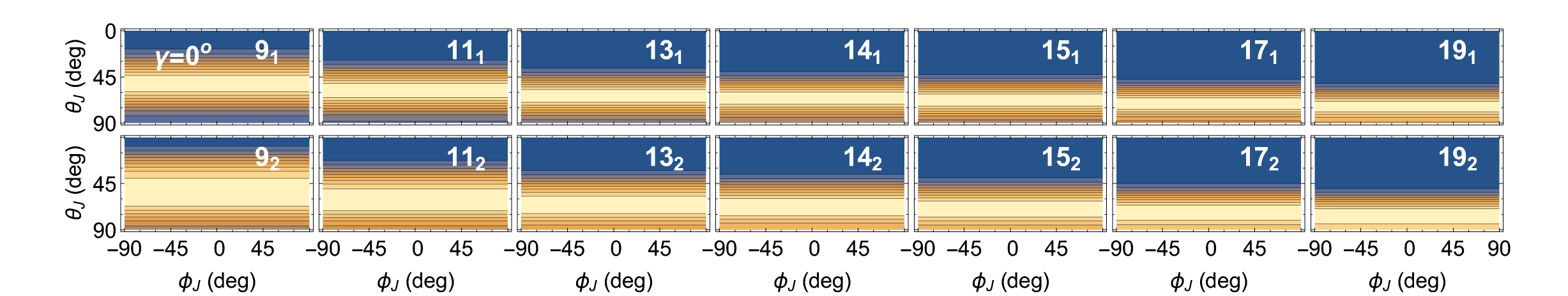}
    \caption{(Color online) The spin coherent state (SCS) maps
    for the yrast (upper panels) and yrare (lower panels) bands 
    with $\gamma$ ranging from $30^\circ$ to $0^\circ$ in steps 
    of $5^\circ$. Only the region $0^\circ \leq \theta\leq 90^\circ $ 
    and $-90^\circ\leq \phi\leq 90^\circ $
    is shown. The other regions are reflection symmetric. Color sequence with increasing
    probability: dark blue-zero level, light blue, dark browns, light brown,
    white.} \label{f:SCS}
  \end{center}
\end{figure*}

\begin{figure*}[!ht]
  \begin{center}
    \includegraphics[width=0.32\linewidth]{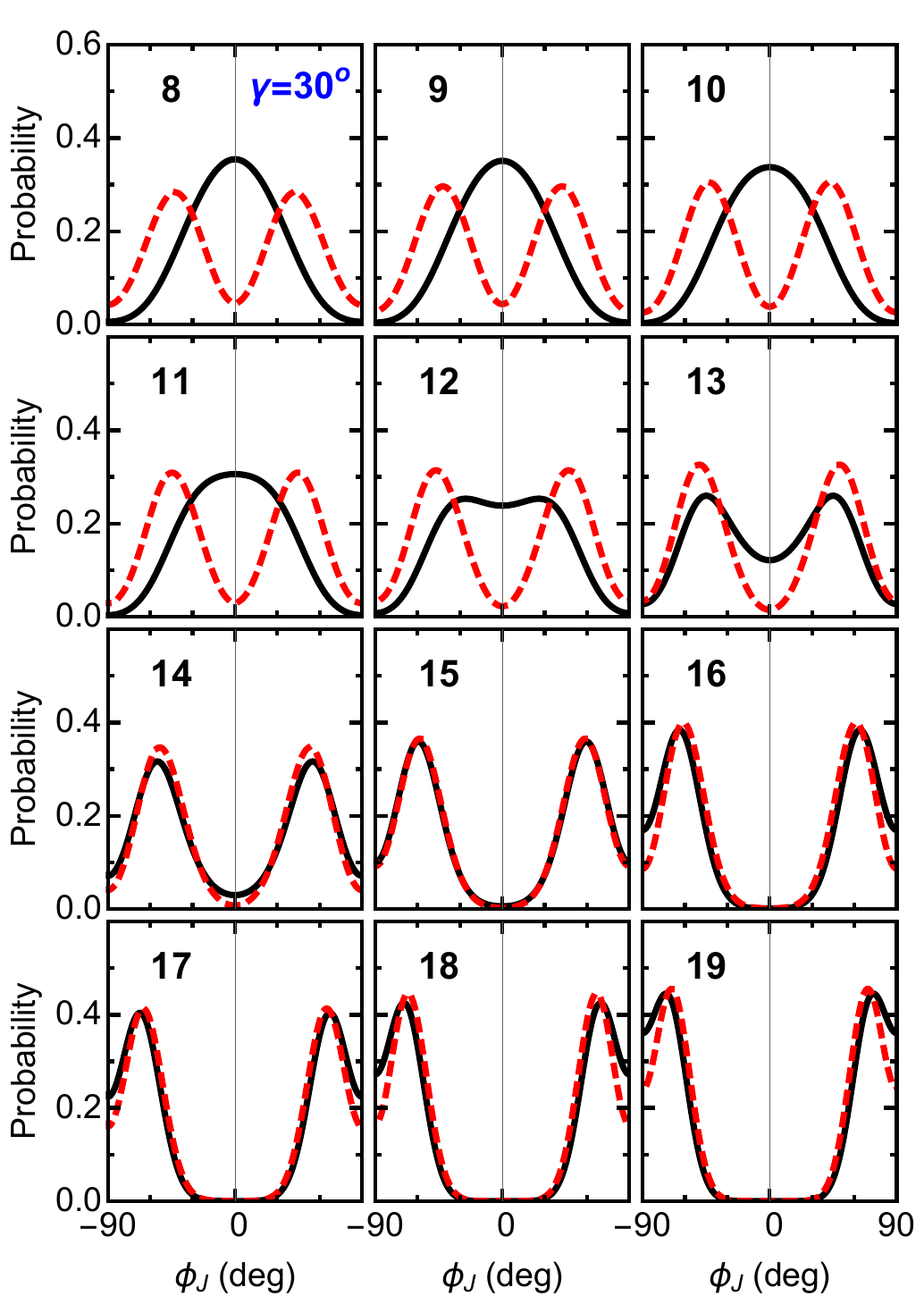}~
    \includegraphics[width=0.32\linewidth]{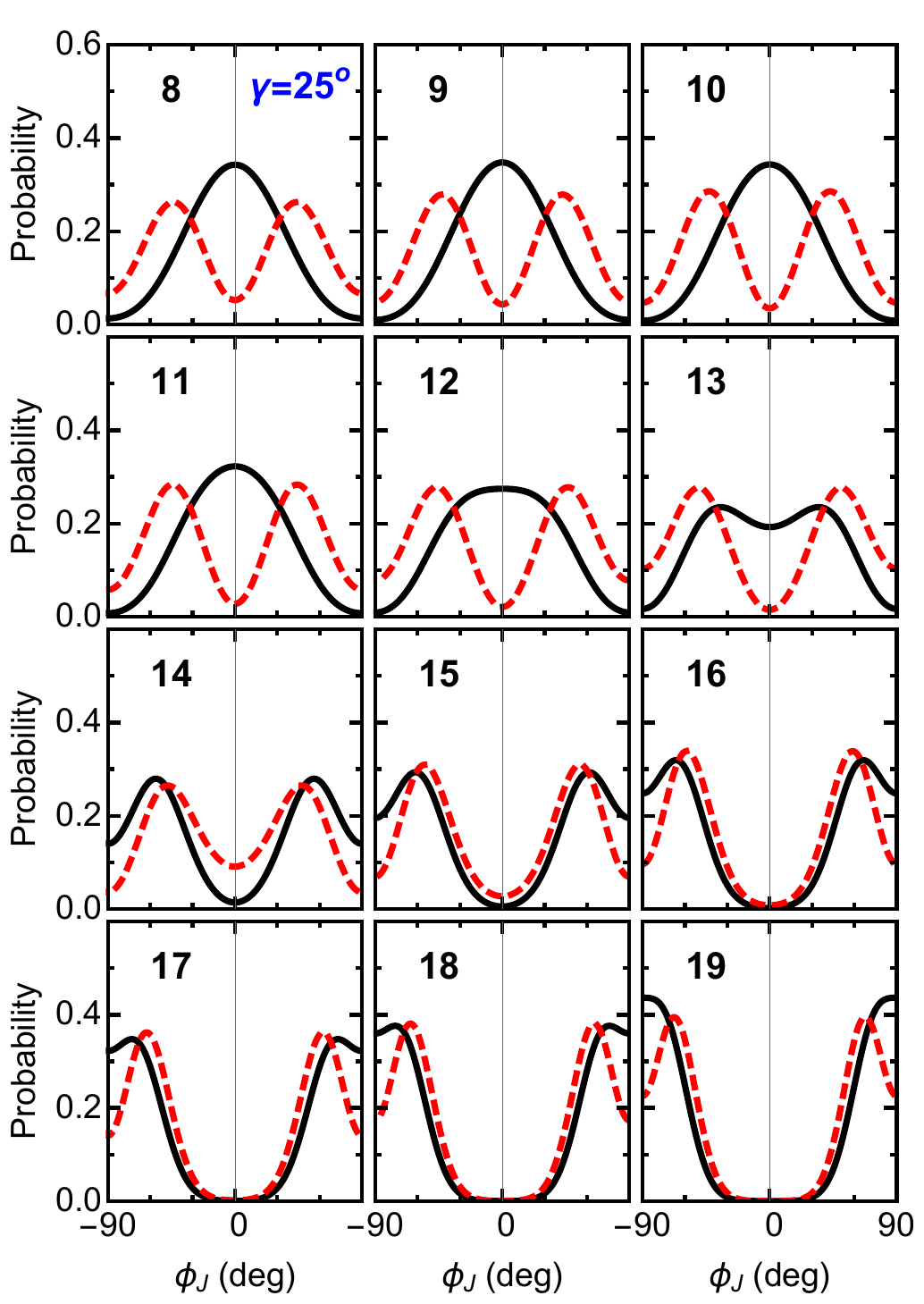}~
    \includegraphics[width=0.32\linewidth]{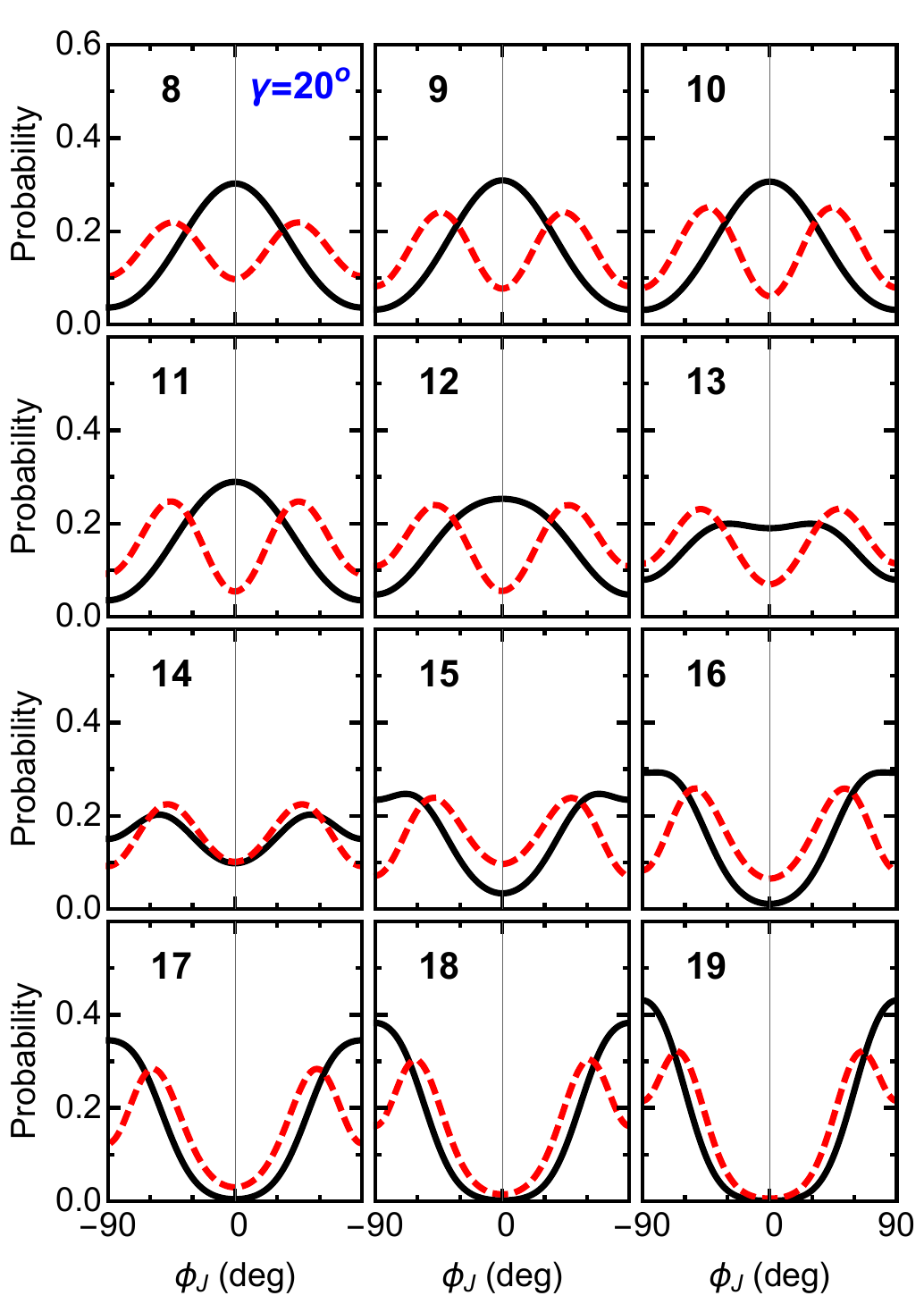}
    \includegraphics[width=0.32\linewidth]{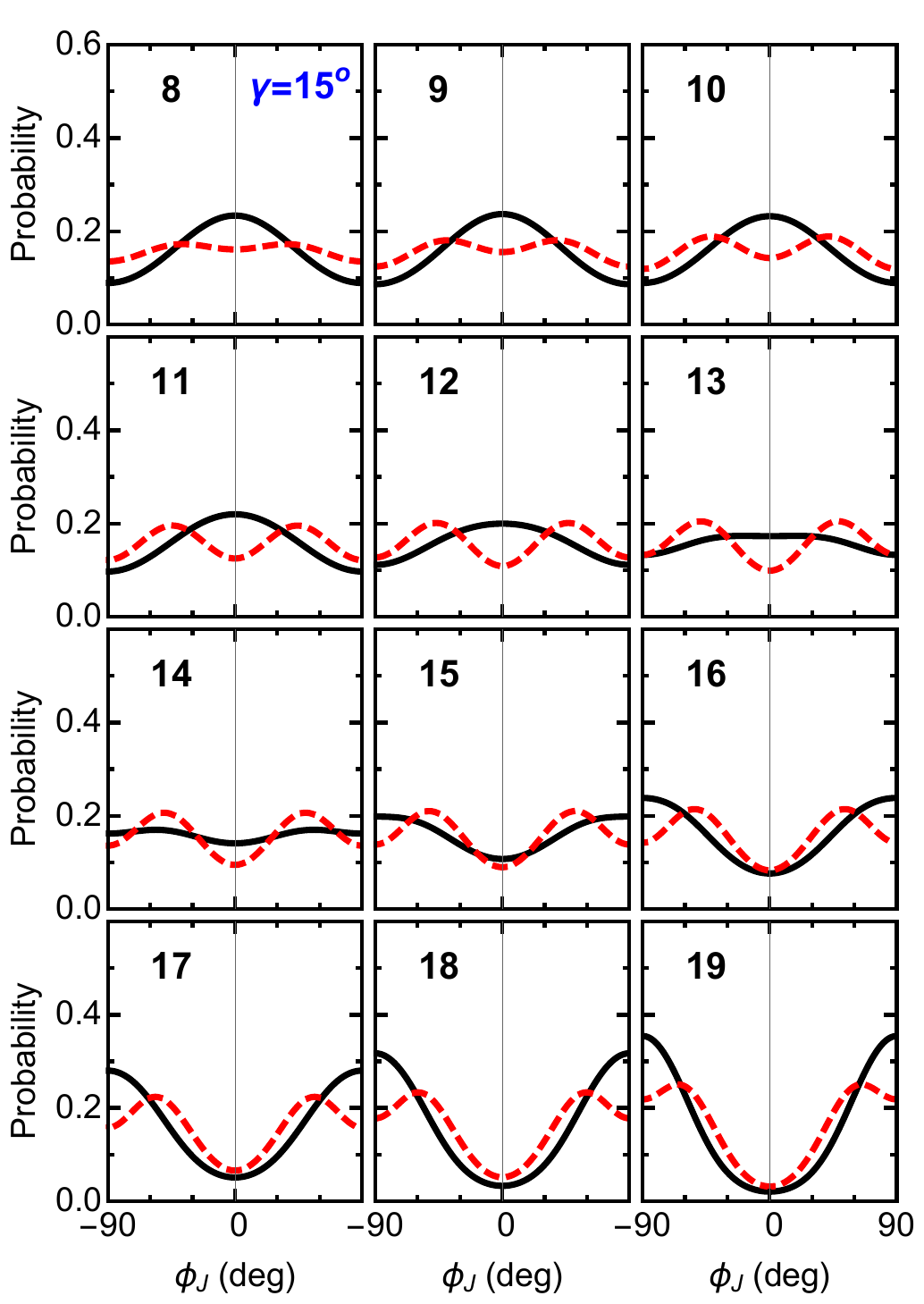}~
    \includegraphics[width=0.32\linewidth]{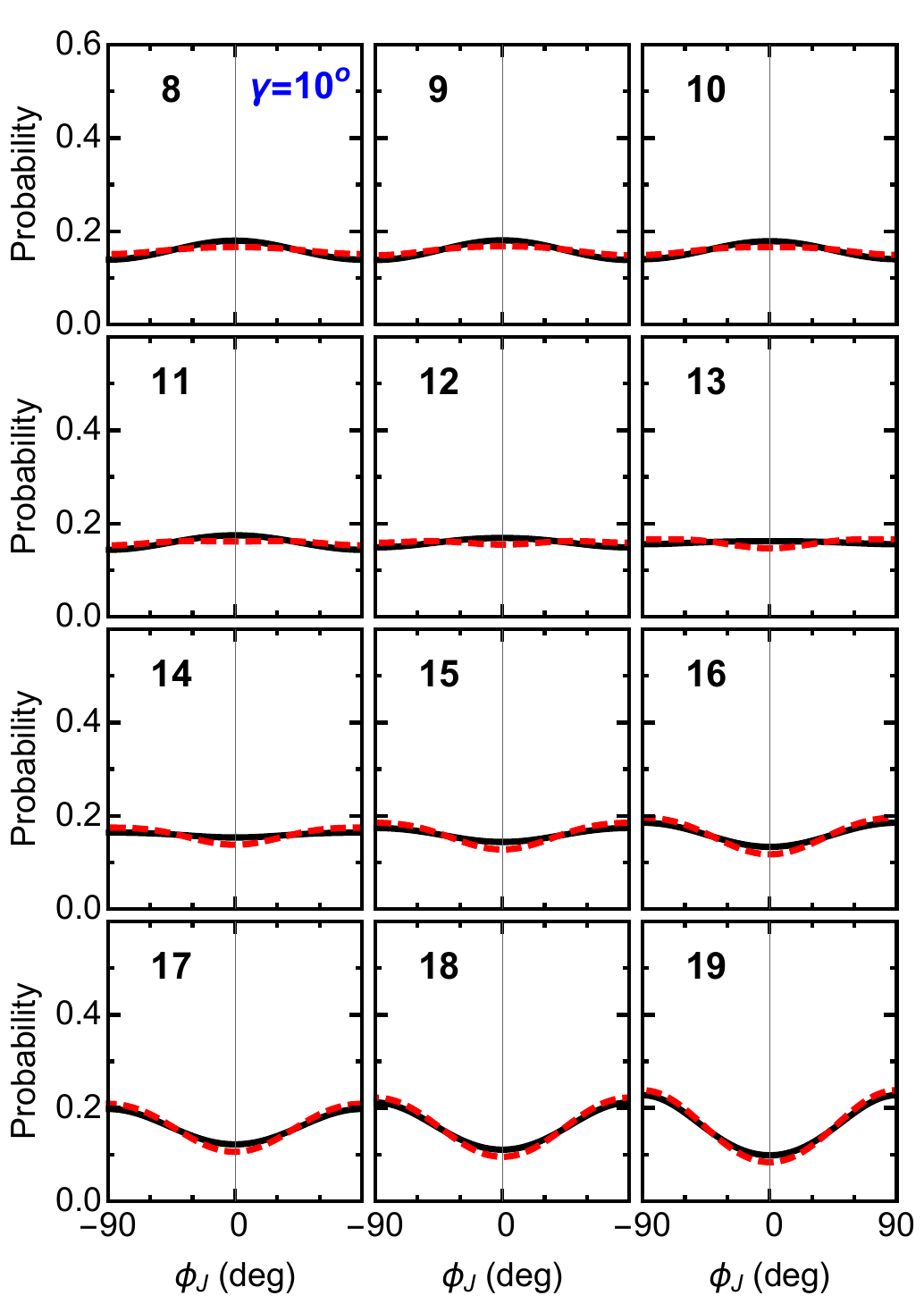}~
    \includegraphics[width=0.32\linewidth]{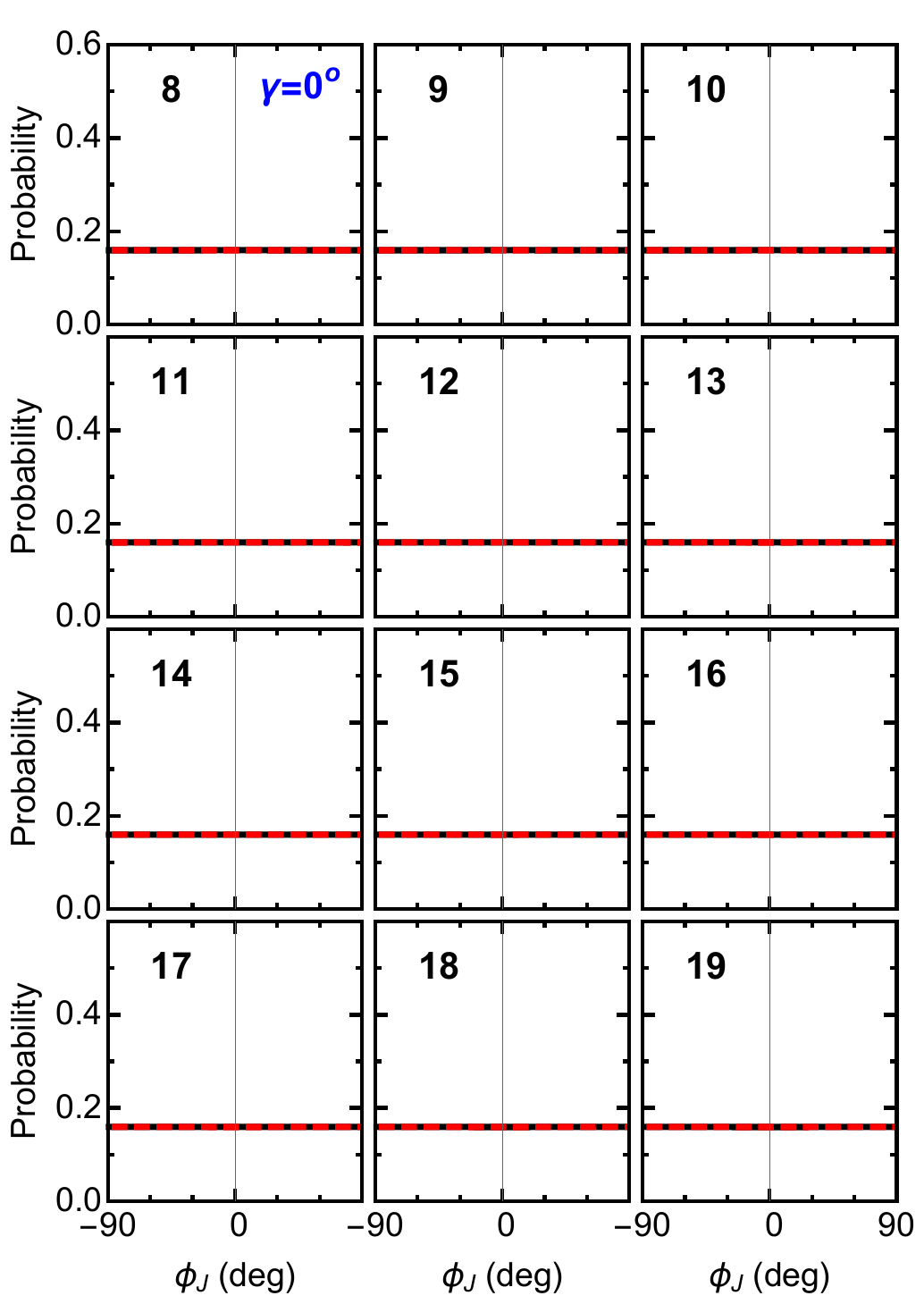}
    \caption{(Color online) The spin squeezed state (SSS) plots 
    for the yrast (solid line) and yrare (dashed line) bands 
    with $\gamma$ ranging from $30^\circ$ to $0^\circ$ in steps 
    of $5^\circ$. }\label{f:SSS}
  \end{center}
\end{figure*}

\subsection{Angular momentum  geometry}

The appearance of the chiral angular momentum geometry~\cite{Frauendorf1997NPA} 
is most intuitively illustrated by the spin coherent state (SCS) maps 
Eq.~(\ref{eq:PSCS})~\cite{Frauendorf2015Conf, Q.B.Chen2022EPJA}, which
 show the probability density $P(\theta, \phi)$ 
 of the angles $\theta$ and $\phi$ of the total angular  
momentum vector with respect to the triaxial shape 
(1: short $s$, 2: medium $m$, 3: long $l$). They are
a proxies of the classical orbits of the total angular 
momentum. The SCS maps  calculated with $\gamma$ ranging 
from $30^\circ$ to $0^\circ$ in step of $5^\circ$ 
are displayed in Fig.~\ref{f:SCS}.

Complimentary visualization of the angular momentum geometry is 
provided by probability density distributions $P(\phi)$, Eq.~(\ref{eq:PSSS}),
of the spin squeezed (SSS) states introduced in Ref.~\cite{Q.B.Chen2024PRC_v1}, 
which show the probability density of the total angular momentum vector 
angle with the $s$ axis.
The azimuthal angle $\phi$ is the
discerning metric for characterizing the degree of chirality 
inherent in the rotational motion~\cite{Frauendorf1997NPA,
Q.B.Chen2013PRC, Q.B.Chen2016PRC}. Specifically, when $\phi=0^\circ$, 
$180^\circ$, $\pm 90^\circ$, the rotational mode manifests 
as a planar rotation, while values of $\phi$ in between imply 
a departure from planar rotation, which indicates the appearance 
of chirality as either CV or CR. In Fig.~\ref{f:SSS} we display 
the SSS probability density $P(\phi)$ calculated 
by Eq.~(\ref{eq:PSSS}). 

\begin{figure*}[t]
  \begin{center}
    \includegraphics[width=0.42\linewidth]{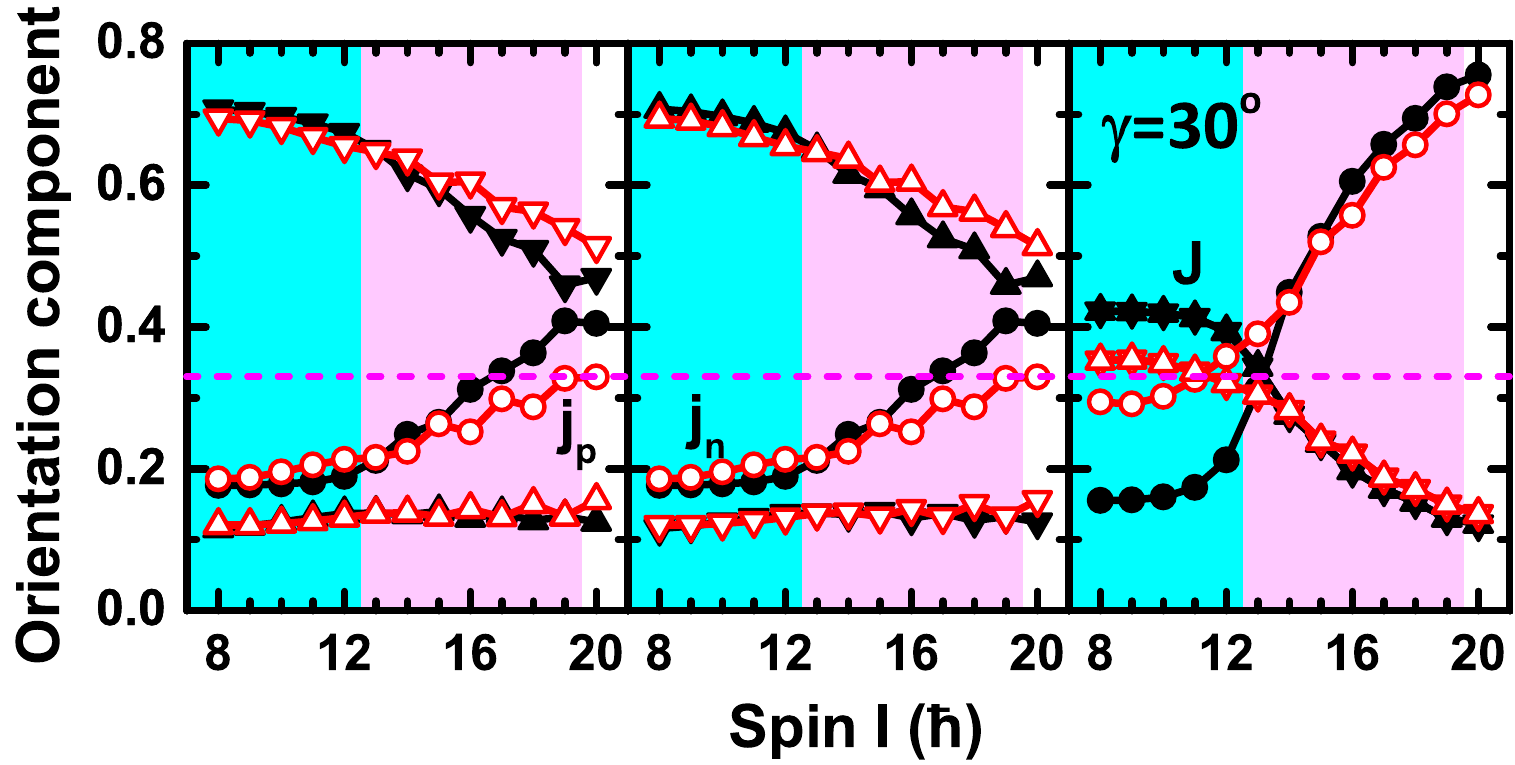}\quad
    \includegraphics[width=0.42\linewidth]{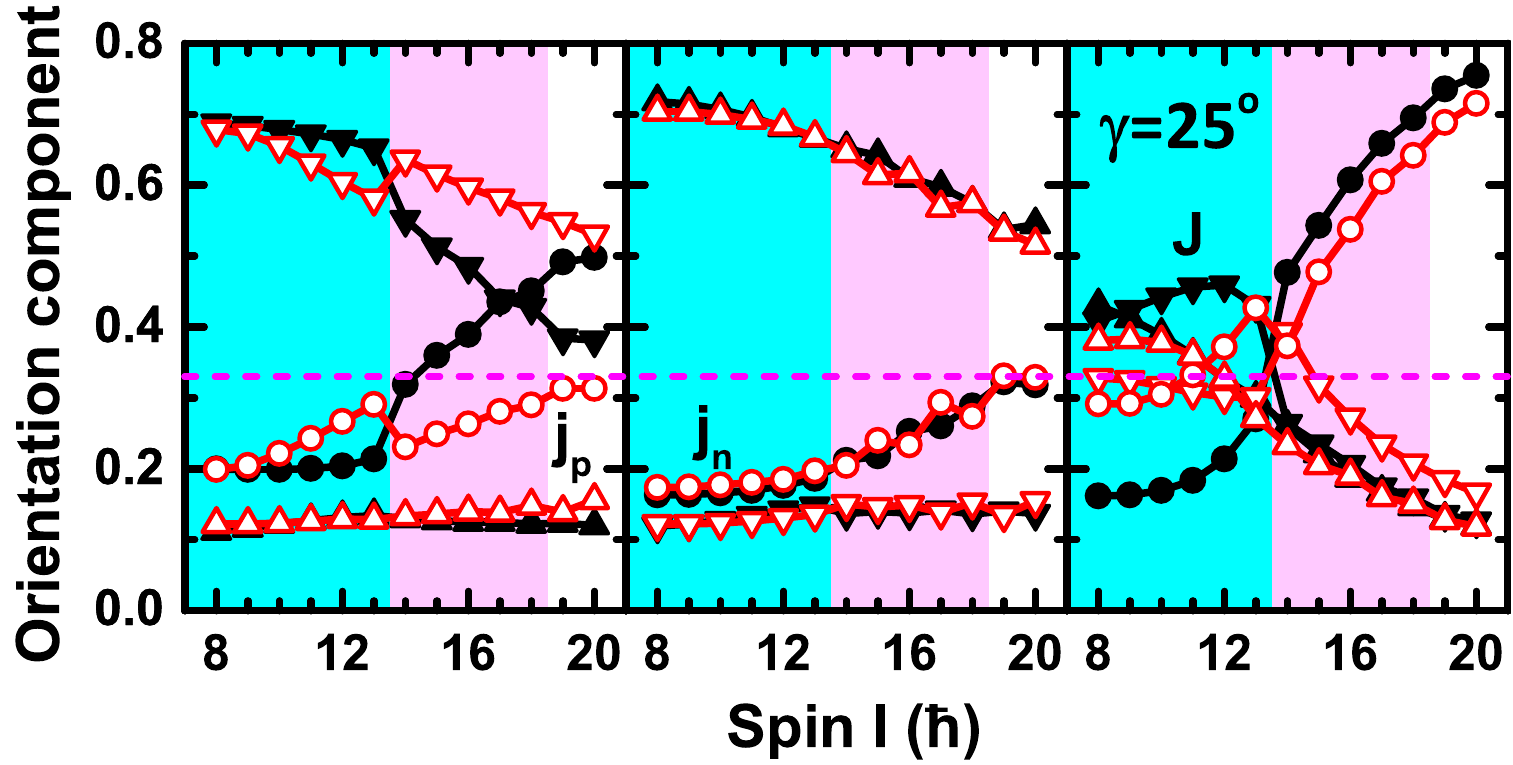}\\
    \includegraphics[width=0.42\linewidth]{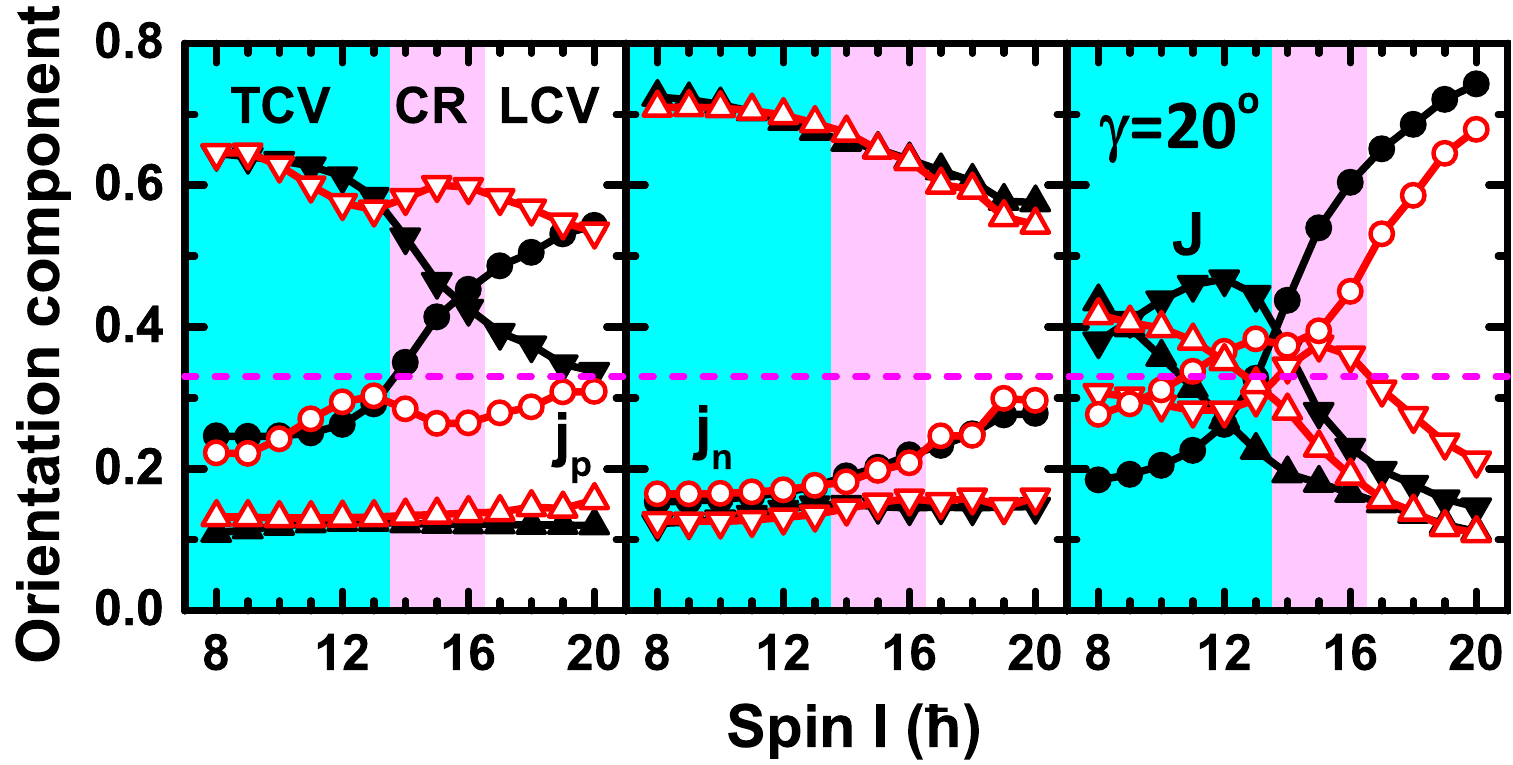}\quad
    \includegraphics[width=0.42\linewidth]{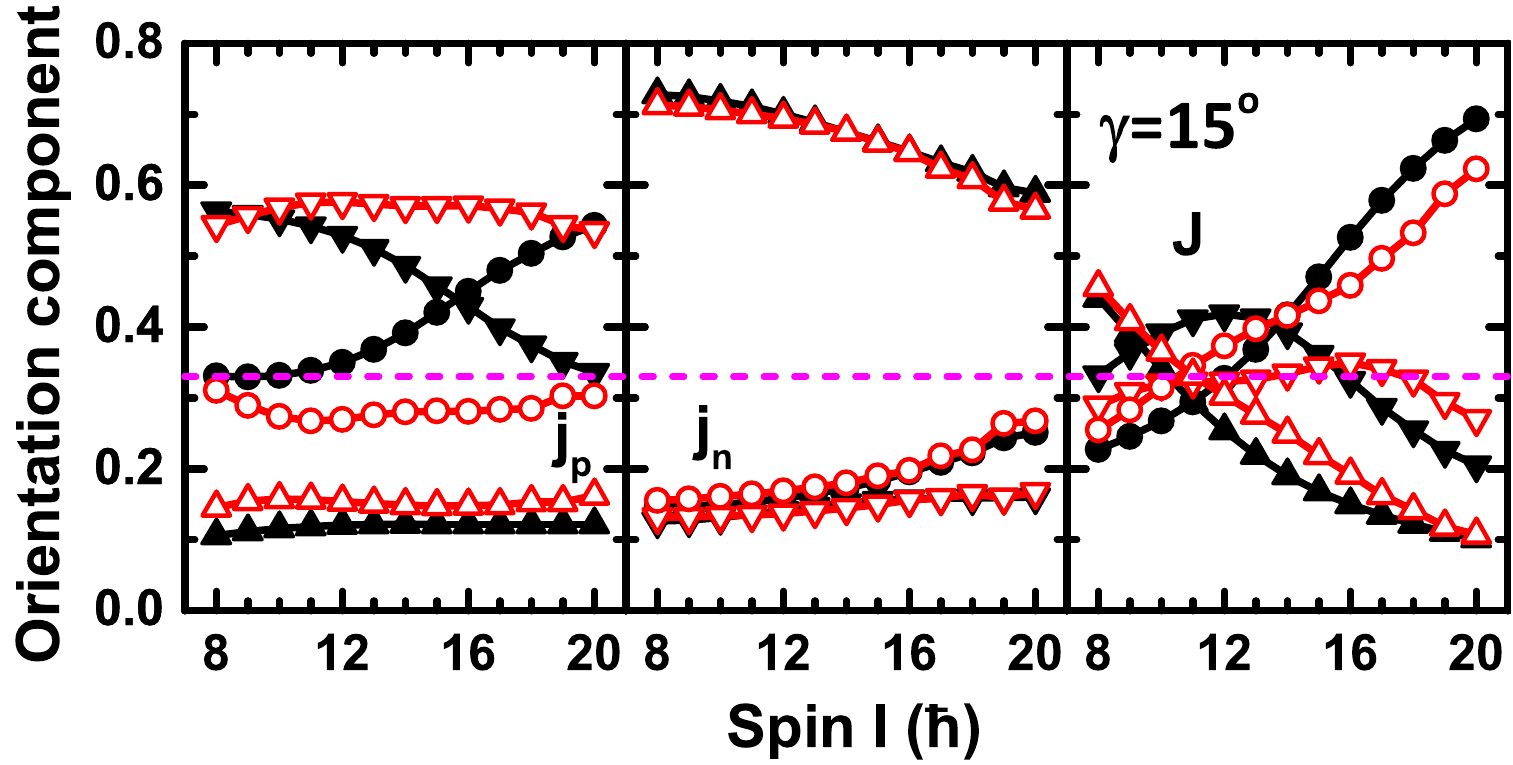}\\
    \includegraphics[width=0.42\linewidth]{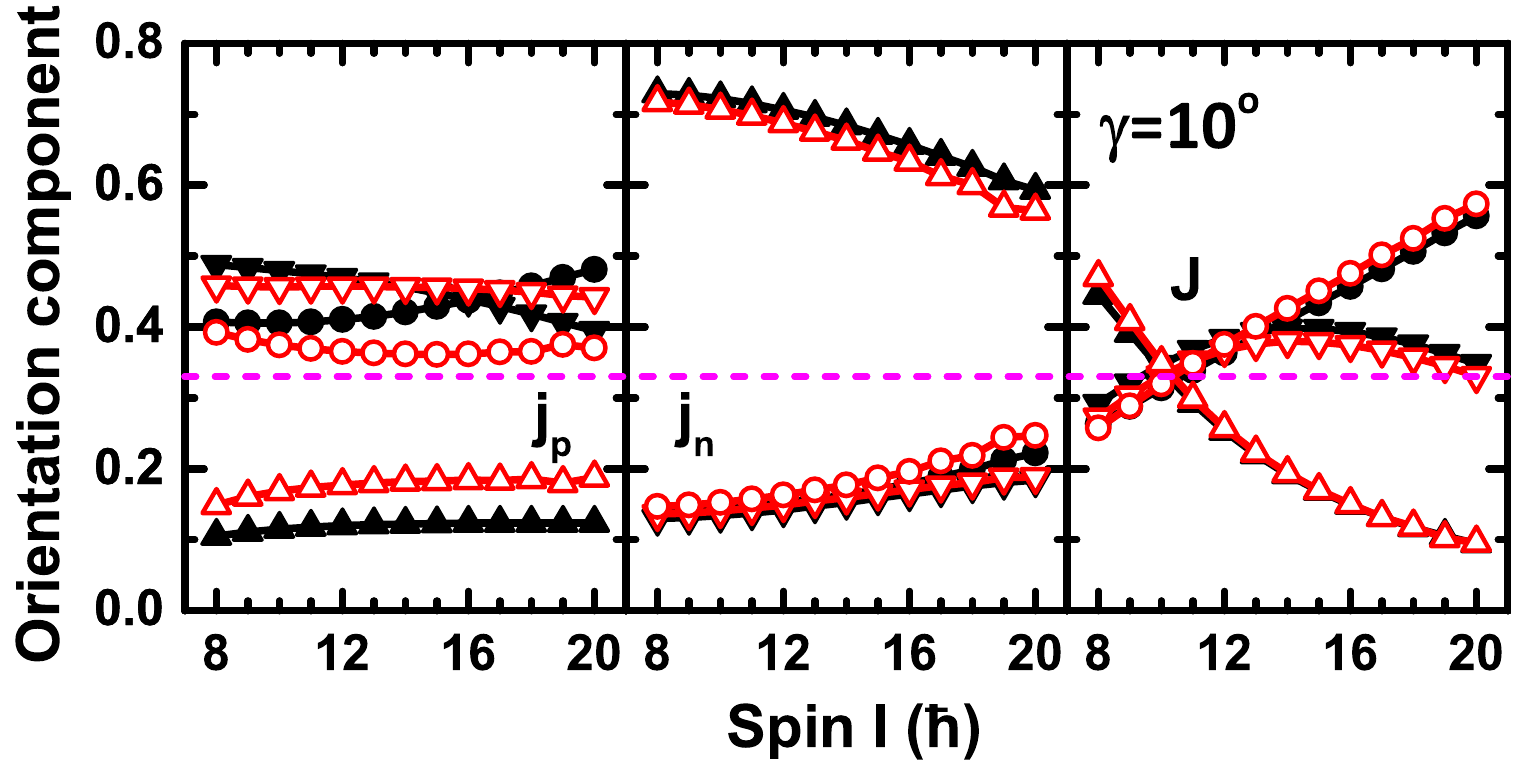}\quad
    \includegraphics[width=0.42\linewidth]{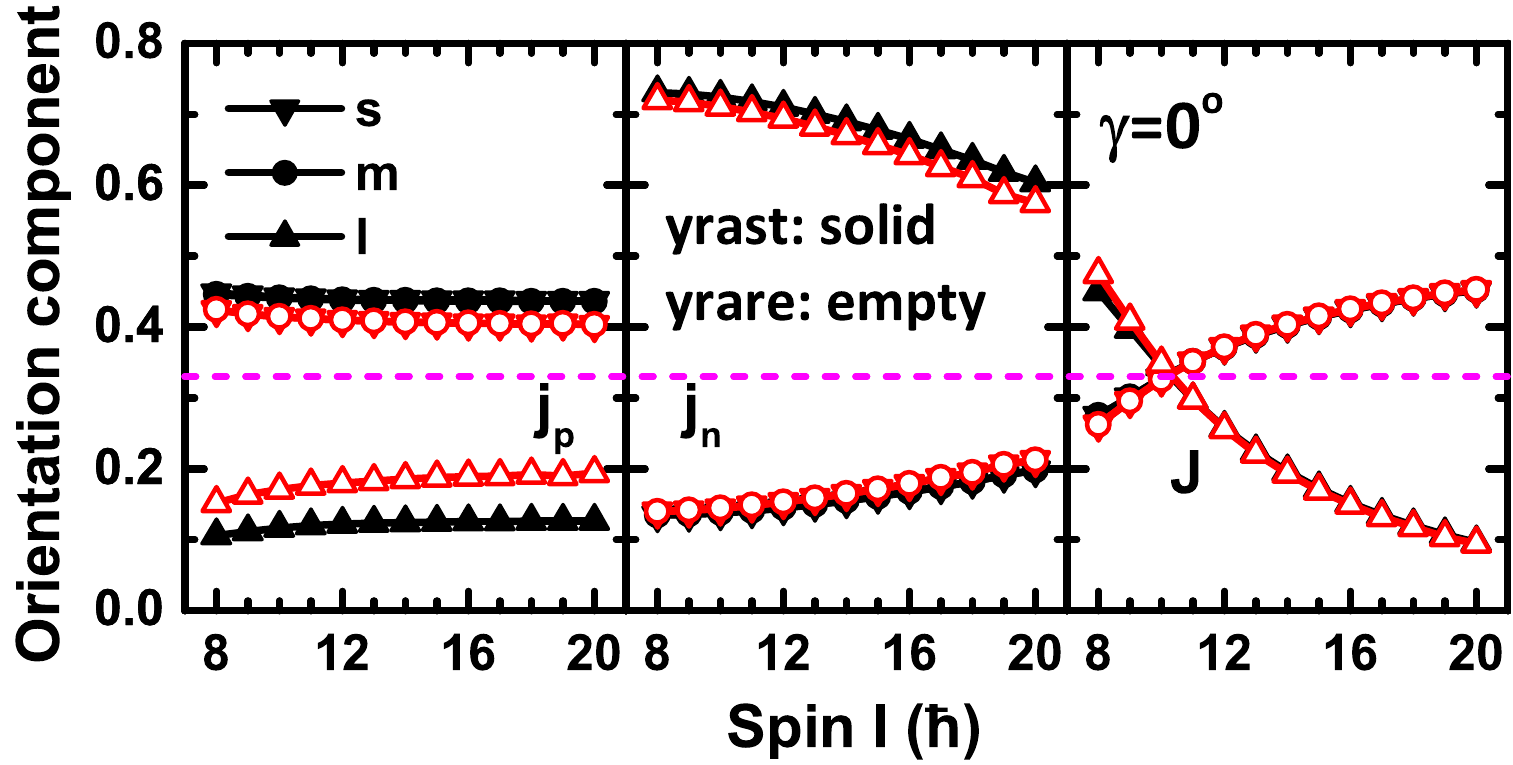}
    \caption{(Color online) The orientation parameters of 
    proton, neutron, and total angular momenta as functions 
    of spin $I$ for the yrast and yrare bands with $\gamma=30^\circ$, 
    $20^\circ$, $15^\circ$, $10^\circ$, and $0^\circ$. The shadows denote the 
    regions of transverse chiral vibration (TCV) and chiral rotation (CR), 
    respectively. }\label{f:Ori}
  \end{center}
\end{figure*}

\begin{figure*}[t]
  \begin{center}
    \includegraphics[width=0.24\linewidth]{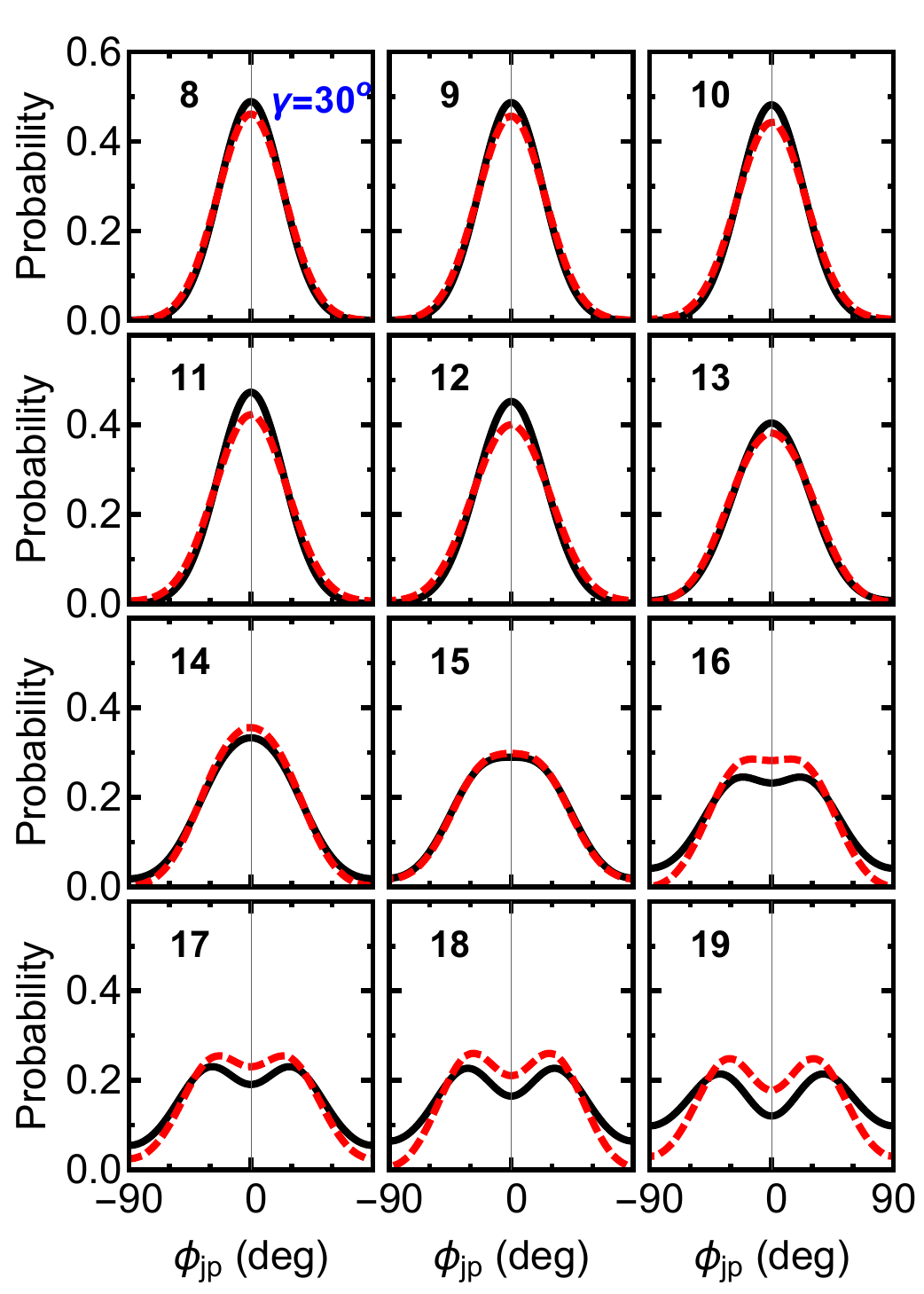}
    \includegraphics[width=0.24\linewidth]{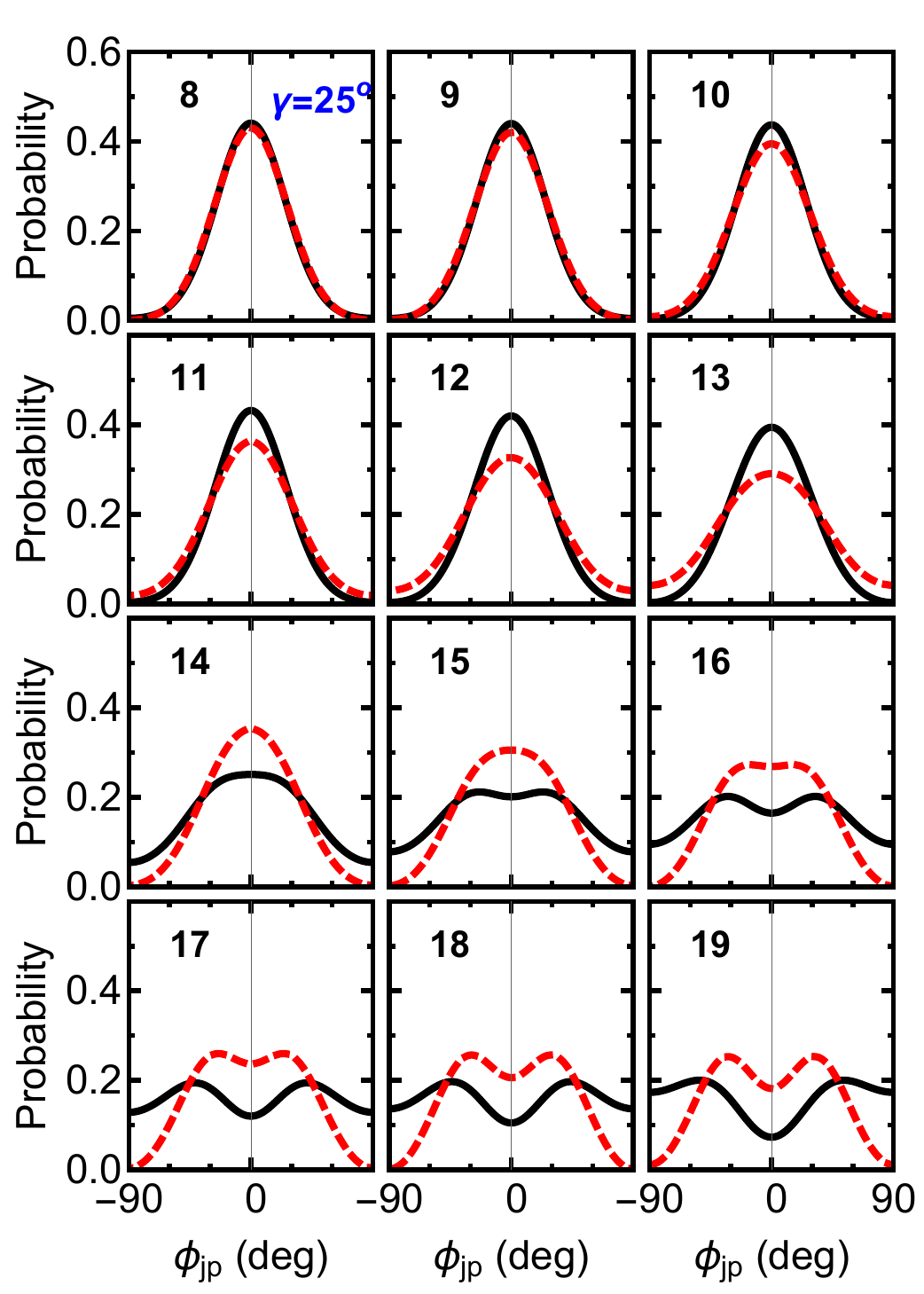}
    \includegraphics[width=0.24\linewidth]{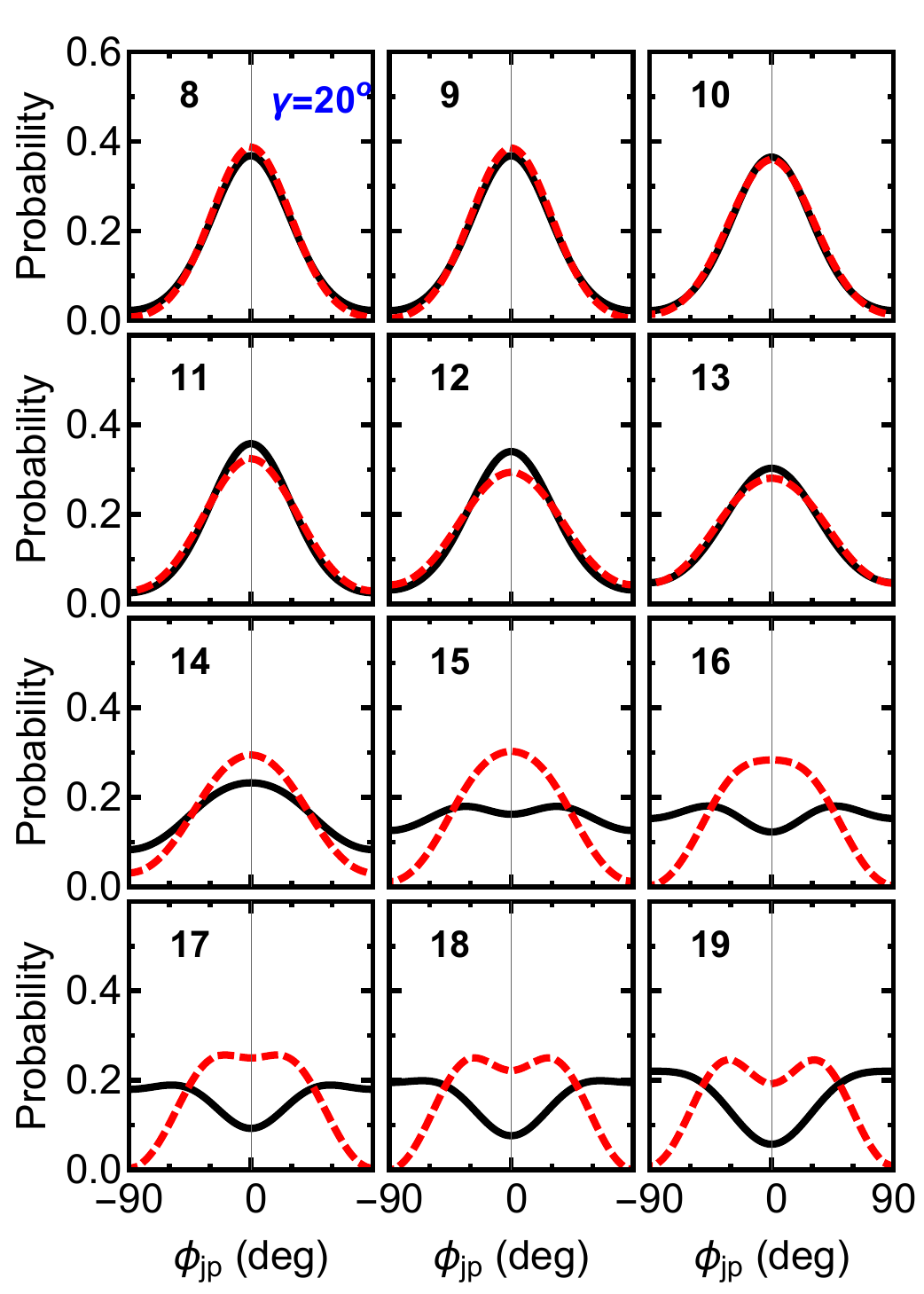}
    \includegraphics[width=0.24\linewidth]{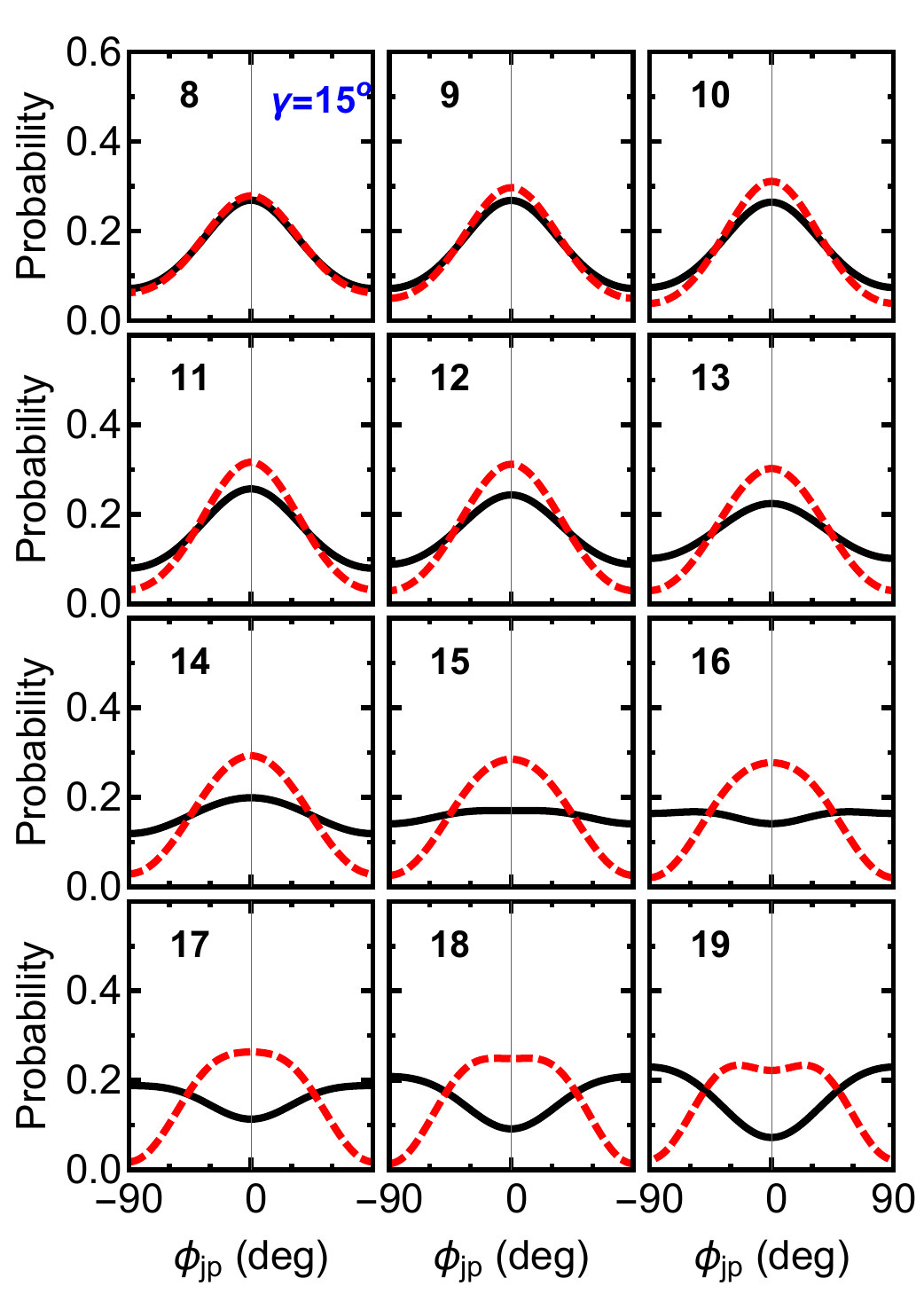}
    \caption{(Color online) The spin squeezed state (SSS) plots 
    for the proton angular momentum in the 
    yrast (solid line) and yrare (dashed line) bands 
    with $\gamma=30^\circ$, $25^\circ$, $20^\circ$, and
    $15^\circ$.}\label{f:SSS_p}
  \end{center}
\end{figure*}

Further complimentary information is provided by the 
orientation parameters (\ref{eq:OriI})-(\ref{eq:Orin}) 
in Fig.~\ref{f:Ori}. They show the squares of the 
orientation angles of the respective angular momenta relative 
to the three principal axes, $o_s=\langle \sin^2\theta\cos^2\phi\rangle$,  
$o_m=\langle \sin^2\theta\sin^2\phi\rangle$, and  
$o_l=\langle \cos^2\theta\rangle$. Large values of
$o_s$, $o_m$, and $o_l$ indicate that the pertaining angular 
momentum vectors are close the $s$, $m$, and $l$ axes, respectively. 
More detailed information about  $\bm{j}_p$ provides Fig.~\ref{f:SSS_p}, 
which shows the SSS probability density $P(\phi_p)$ of its angle 
with the $s$ axis. 

The most favorite conditions  for chirality appear for maximal traxiality of $\gamma=30^\circ$.

For low $I$, the yrast state is a blob centered at $\theta=45^\circ$ 
and $\phi=0^\circ$ in the SCS map (cf. Fig.~\ref{f:SCS}) and a maximum 
at $\phi=0^\circ$ in the SSS plot (cf. Fig.~\ref{f:SSS}), which correspond 
to uniform rotation about the axis tilted by $45^\circ$ into 
the $s$-$l$ plane. The planar geometry is recognized by 
the small angle $\phi$ of $\bm{J}$ in Fig.~\ref{f:Ori}. 

The yrare state has a minimum at the tilted axis with a 
density rim around it, which depicts the  wobbling 
of the total angular momentum $\bm{J}$ around the tilted axis. 
Correspondingly, in the SSS plot there is a minimum at $\phi=0^\circ$ 
and two maxima at $\phi=\pm 45^\circ$. This characterizes 
the yrare band as a vibrational excitation, which 
accordingly, is called as ``chiral vibration" (CV). Since the total 
angular momentum oscillates with respect to the $s$-$l$ plane, 
which is perpendicular to the $m$ axis with the largest moment 
of inertia, we call this motion transverse chiral vibration (TCV). 
Its experimental signature is similar to the transverse 
wobbling (TW)~\cite{Frauendorf2014PRC}, which is decreasing 
energy splitting $\Delta E(I)$ between the doublet bands as shown 
in Fig.~\ref{f:DeltaE}. Like the TW mode, the TCV becomes 
unstable toward CR, which is indicated by the appearance of a dip
of the yrast probability at $\phi=0^\circ$.

Around $I=15$ the angular momentum is localized near $\theta=65^\circ$, 
$115^\circ$ and $\phi=\pm 60^\circ$ (and the four equivalent 
points on the backside of the hemisphere) for both 
the yrast and the yrare states. At each of these points, 
the $\bm{J}$, $\bm{j}_p$, and $\bm{j}_n$ arrange as 
a left-handed and a right-handed vector triples with 
the same energy, where adjacent octants harbor states 
of opposite handedness 

That is, the chiral symmetry is spontaneously broken. The 
symmetry-broken states combine into two states with restored chiral 
symmetry, the yrast and yrare states, which have nearly the same energy. 
Their splitting $\Delta E(I)$ reflects the couplings between the octants 
of chiral states. The probabilities of the SSS states $P(\phi)$ 
are nearly identical. This region has been called as the ``chiral rotation" (CR).
It is also referred to as aplanar rotation because, as seen 
in Fig.~\ref{f:Ori}, the proton and neutron angular momenta $\bm{j}_p$
and $\bm{j}_n$ are near the $l$-$s$ plane while the total 
angular momentum $\bm{J}$ is substantially out of this plane.
We define the CR region by the condition that the yrast state must have 
a maximum located out of the planes $\phi=0^\circ$, $\phi=90^\circ$ 
and $\theta=90^\circ$.  

Above $I=15$,  the spatial distribution  of the yrast and yrare states
remains similar while the four blobs on the front hemisphere move toward  
$\theta = 90^\circ$ and $\phi = \pm 90^\circ$ with increasing $I$. It retains the
chiral geometry while the coupling between the states of opposite 
chirality increases, which is  seen best in Fig.~\ref{f:SSS}.

For $I=19$, the SCS probability of the yrast state has its maxima at 
$\phi=\pm 90^\circ$, which indicates the boundary of the CR region.
The SSS probability $P(\phi)$ of the yrast state has  
a small dip at $\phi=\pm 90^\circ$, while $P(\phi)$ of the yrare state 
is substantially smaller. These are characteristics for states carrying 
respectively zero or one quantum of an inharmonic vibration  
with respect to the $m$ axis. Since the total angular momentum 
aligns along the $m$ axis, which corresponds to the principal 
axis of inertia with the largest moment of inertia, this
motion is referred to as longitudinal chiral vibration (LCV). 
Its experimental signature is analogous to that 
of longitudinal wobbling (LW)~\cite{Frauendorf2014PRC}, 
manifested as an increasing energy splitting between the 
doublet bands as illustrated in Fig.~\ref{f:DeltaE}.
Thus, the region of LCV has been encountered. 

The energy difference $\Delta E$ and structural difference between the
two bands result from superposition 
of the PTR  states of opposite  chirality and planar configurations 
in order to restore chiral symmetry~\cite{Starosta2017PS, Q.B.Chen2018PRC}. 
Chiral symmetry of the PTR Hamiltonian guarantees that each left-handed 
component has a right-handed partner with the same energy that is 
generated by the chirality-reversing operation, where the planar
components do not have such partners.
In SCS representation the planar configurations are located in the planes 
$\phi=0^\circ,~180^\circ$, where $J_m=0$, the planes $\pm 90^\circ$,
where $(j_p)_m=\pm 1/2$, and the planes $\theta=90^\circ$, where $(j_n)_l=\pm 1/2$.
Chirality invariance is restored by forming even or odd superpositions. 
At the TCV-CR  boundary the states at  the $\phi=0^\circ,~180^\circ$ 
plane dominate. The even combination includes the planar components 
while the odd combination does not~\cite{Q.B.Chen2018PRC}, 
which is seen as $P(\phi=0^\circ)=0$ in Figs.~\ref{f:SCS} and \ref{f:SSS}. 
At the CR-LCV boundary the states at the $\phi=\pm 90^\circ$
plane dominate. Here the ``planar" states carry 1/2 units 
of angular momentum perpendicular to the plane. These terms 
appear with opposite sign in the even and odd linear combinations, 
which result in a respectively large and small values of $P(\phi=\pm 90^\circ)$.

As seen in Fig.~\ref{f:SCS}, for $I=19$ 
the SCS probability $P(\theta=90^\circ,\phi)$ at its maximum 
is larger for the yrast than the yrare states, which is 
reflected by the staggering of $E(I)$ in Fig.~\ref{f:Energy}.
The staggering of the  energies with $I$  is generated by the alternating 
sign of the matrix elements $j_\bot$ between the ``planar" states that 
carry 1/2 units of angular momentum perpendicular to $\theta=90^\circ$ plane 
in the expression for the rotational energy, which groups the members
of the rotational band according to their ``signature"~\cite{Bohr1975}.

The orientation parameters for the proton in Fig.~\ref{f:Ori}
are similar for the yrast and yrare states, where $o_l^p \approx 0.1$
indicates that $\bm{j}_p$ is located in the $s$-$m$ plane. For $I=8$
it is near the $s$ axis. With increasing $I$, it is tilted into the $s$-$l$
plane reaching an average  angle of $\approx 43^\circ$ for the yrast state
and $\approx 39^\circ$ for the yrare state  at $I=19$. The proton SSS distributions
in Fig.~\ref{f:SSS_p} reflect the slight difference in the orientation parameters 
for the largest $I$, which is caused by the admixture of the planar 
states in the $\phi=\pm 90^\circ$ plane. For symmetry reasons, 
the $\bm{j}_n$ is tilted in the same way into the $l$-$m$ plane, 
as seen in the neutron panel of Fig.~\ref{f:Ori}.

As illustrated in Figs.~\ref{f:SCS} and \ref{f:SSS},
the cases of $\gamma = 25^\circ$ and $20^\circ$ show   
similar topologies of TCV, CR, and LCV as $\gamma = 30^\circ$. 
With decreasing $\gamma$, the probability densities of the CR yrast 
states increase in the planar regions  
that connect the locations of opposite chirality ($\phi=0^\circ$, $\pm 90^\circ$),
which lowers their energy. This 
is reflected by the increase  of $\Delta E(I)$ in Fig.~\ref{f:DeltaE}, 
because the probability distributions of the yrare states do not 
change much. Additionally, the critical spin $I_c$, at which 
the two peaks emerge in the yrast band, increases. This behavior 
can be attributed to decreases in the ratios of $\mathcal{J}_m/\mathcal{J}_s$ 
and $\mathcal{J}_m/\mathcal{J}_l$. 

The SCS probability density $P(\theta=90^\circ,\phi)$  is smaller for 
$\gamma=25^\circ$ than for $\gamma=30^\circ$ and yet smaller
for $\gamma=20^\circ$, which is reflected by the disappearance 
of the staggering in Figs.~\ref{f:Energy} and \ref{f:DeltaE}.

Figure~\ref{f:Ori} shows that weaker triaxiality of the potential results into
a more rapid tilting of $\bm{j}_p$ into the $s$-$m$ plane. 
At  $I=16$,  the center of the CR region, its average angle with the $s$ axis is
$\approx 42^\circ$ for the yrast state and $\approx 35^\circ$ for the yrare state
in case of $\gamma=25^\circ$ and, respectively, $45^\circ$ and $33^\circ$
in case of $\gamma=20^\circ$. These values are consistent with 
the proton SSS probability densities in Fig.~\ref{f:SSS_p}. With 
decreasing $\gamma$, the planar states  become more important. 
The functions $P(\phi_p)$ of the yrast and yrare states differ 
most for $\phi=0^\circ,~\pm 90^\circ$, while they stay close 
together around $\phi=\pm 45^\circ$.

The $I$ dependence of the structure of the yrast and yrare states 
suddenly interchanges between $I=13$ and 14. The yrast states 
continue the structural development of the yrare states and
vice versa. It is clearly seen for the  orientation parameters
of $\bm{j_p}$ and $\bm{J}$ (Fig.~\ref{f:Ori}), the SSS probability
densities $P(\phi_p)$ (Fig.~\ref{f:SSS_p}) and $P(\phi)$ 
(Fig.~\ref{f:SSS}), as well as the SCS maps $P(\theta,\phi)$ 
(Fig.~\ref{f:SCS}). The structural continuity is also obvious 
in the energies $E(I)$ (Fig.~\ref{f:Energy}). 

For $\gamma=20^\circ$ the coupling to the planar states is yet stronger.
In Figs.~\ref{f:DeltaE} and \ref{f:Ori} the crossing between the bands 
changed into an avoided crossing. The SCS and SSS plots for $\bm{J}$ 
are further smoothed out, but still quite similar for $I=14$, 
which indicates chiral structure. The  SSS probability densities 
for the proton $P(\phi_p)$ in Fig.~\ref{f:SSS_p}
are still reasonably  similar for the yrast and yrare states. 

The restoration of the chiral invariance in the narrow  
CR region around $I=14$ causes a certain reorientation 
of the odd proton. It becomes stronger in the LCV regime 
when the $\bm{J}$ approaches the $\phi=\pm 90^\circ$ plane with
increasing $I$. In general, the difference between 
the orientation of the proton in the yrast and yrare states 
increases with decreasing triaxiality, which is expected.  

The $\gamma=0^\circ$ panels illustrate the axial symmetric limit. 
The SCS and SSS probabilities of the states do not depend on $\phi$ with  
$P(\phi)=1/2\pi \approx 0.159$. Accordingly, 
$o_s=o_m$, because $\mathcal{J}_m=\mathcal{J}_s$ and $\mathcal{J}_l=0$. 
The yrare states are excited from the yrast states  by changing  the 
orientation of $\bm{j}_p$ and $\bm{j}_n$ with respect to the $l$ axis, 
which is  seen in Fig.~\ref{f:Ori} and in $P(\theta, \phi)$ of the SCS maps  
Fig.~\ref{f:SCS}. The spin increases by adding rotor angular momentum 
$\bm{R}$ perpendicular to the symmetry axis with equal probability 
$P(\phi)/2\pi$. In the axial potential $P(\phi)=1/2\pi\approx 0.159$ 
for the quasiparticle angular momenta as well. Due to the Coriolis 
interaction $\bm{j}_p$, $\bm{j}_n$, and $\bm{J}$ arrange in a plane.

For $\gamma = 10^\circ$, the yrare states still represent excitations 
by reorienting the quasiparticle orientations while 
the triaxiality becomes noticeable. For low $I$, the SCS and SSS
probabilities are slightly enhanced at $\phi=0^\circ$ because 
$\bm{j}_p$ prefers this orientation. With increasing  $I$ 
the probability maxima move to $\phi=\pm 90^\circ$ because
$\mathcal{J}_m >\mathcal{J}_s$, which eventually prevails.
The relocation of the maxima is reflected by the drop of $o_s^I$
for $\bm{J}$ in the right panel of Fig.~\ref{f:Ori}. 

The $\gamma = 15^\circ$ case has transitional nature.  
The SCS and SSS plots may be seen as the strongly washed out 
pattern of $\gamma = 20^\circ$. While the double peak 
structure of the yrare states survives in smoothed form, the 
maxima of the yrast states directly change from $\phi=0^\circ$ 
to $\pm 90^\circ$ without the intermediate double hump 
that generates the CR region, which is missing. Alternatively, 
SCS and SSS plots may be seen as a magnification of the differences 
between the $\gamma = 10^\circ$ and $0^\circ$.

\subsection{Concurrence triangle}

\begin{figure}[t]
  \begin{center}
    \includegraphics[width=0.95\linewidth]{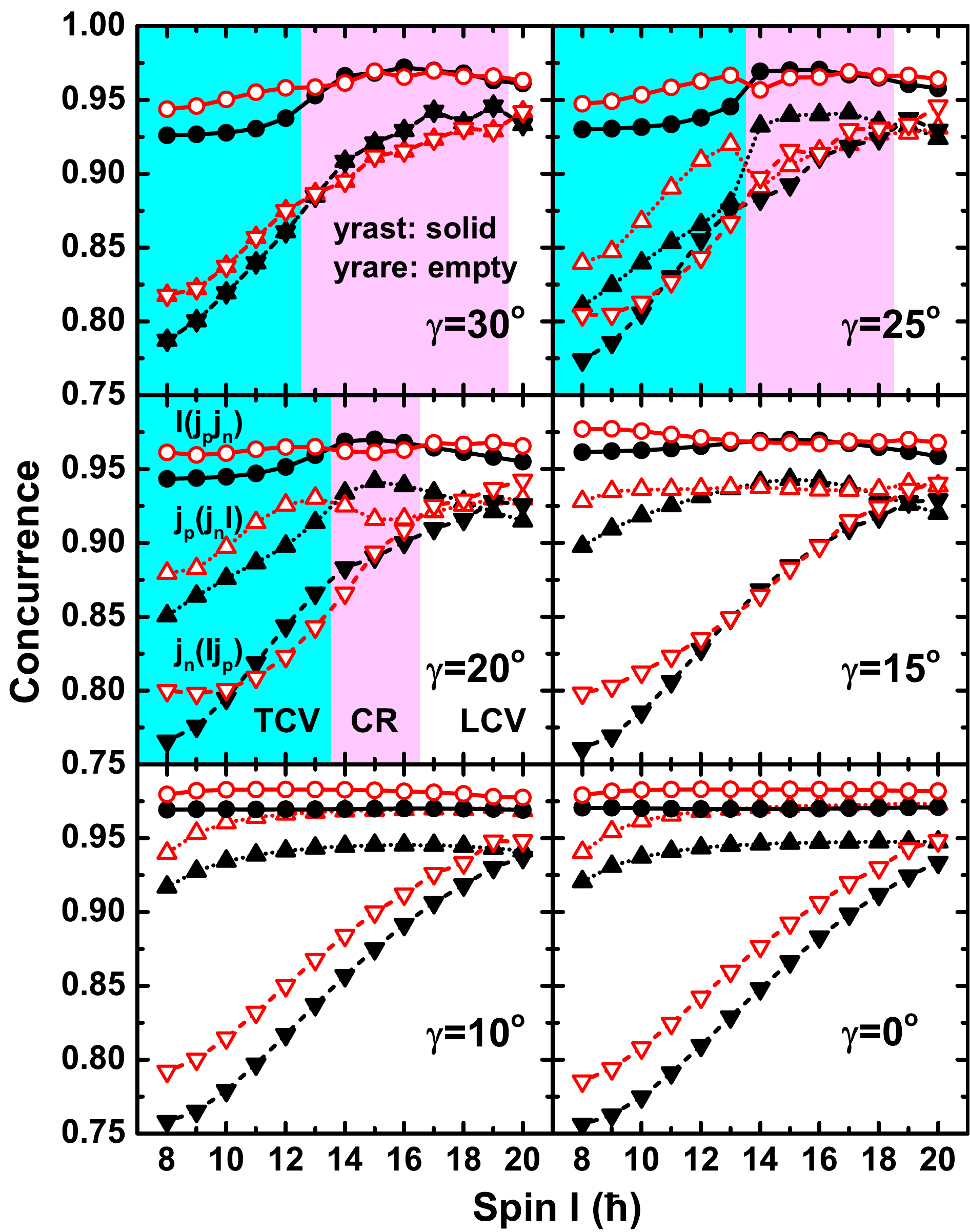}
    \caption{(Color online) The calculated lengths of concurrence triangle
    $\mathcal{C}_{I(j_pj_n)}$, $\mathcal{C}_{j_p(j_n I)}$, and 
    $\mathcal{C}_{j_n(Ij_p)}$ as functions of spin $I$ for the 
   yrast and yrare bands with $\gamma$ ranging from $30^\circ$ 
    to $0^\circ$ in step of $5^\circ$. The shadows denote 
    the regions of transverse chiral vibration (TCV)
    and rotation (CR), respectively.}\label{f:Concurrence}
  \end{center}
\end{figure}

Figure~\ref{f:Concurrence} shows the lengths of concurrence 
triangle sides, $\mathcal{C}_{I(j_pj_n)}$, $\mathcal{C}_{j_p(j_nI)}$, 
and $\mathcal{C}_{j_n(Ij_p)}$, calculated from the 
eigenvalues of the reduced density matrix as 
outlined in Eqs.~(\ref{eq:ConI})-(\ref{eq:Conjn}), which
respectively correspond to the bipartitions $I(j_p j_n)$, 
$j_p(j_n I)$, and $j_n(Ij_p)$. 

The concurrence $\mathcal{C}_{I(j_pj_n)}$ measures the entanglement of 
the rotor and the proton particle and the neutron hole (purity measure introduced 
in Ref.~\cite{Q.B.Chen2024PRC_v2}). The two subsystems are strongly entangled 
right from $I=8$. The rotor on its own would wobble about the $m$ axis 
with the largest moment of inertia. Its interaction with particle and 
hole reorients $\bm{J}$ such that it wobbles about the 
$\bm{j}_p+\bm{j}_n$ axis (TCV).

For $\gamma = 30^\circ$, that ratio $o_m/o_s$ corresponds to an 
average angle of $31^\circ$. The maximal $\mathcal{C}_{I(j_pj_n)}$ 
is observed around $I = 15$, where  chirality is most favored.
This behavior can be understood in the context of 
chirality, which requires the three components of angular momentum 
along the principal axes to be comparable in magnitude. In the 
high spin region, where the LCV is encountered, 
the $\mathcal{C}_{I(j_pj_n)}$ decreases. The decrease of the entanglement
can be attributed to the approach of wobbling about $m$ axis, which 
is the mode of the uncoupled rotor. For the yrast states the orientation 
parameters correspond to average angles $69^\circ$ of $\bm{J}$ with 
the $m$ and $l$ axes.

Such behavior is reminiscent of the evolution of the TW
mode studied in Ref.~\cite{Q.B.Chen2024PRC_v2} 
for one proton coupled to the rotor. Specifically, 
as the TW mode collapses, the entropy 
increases, whereas the establishment of the LW mode 
leads to a reduction in entropy.

For $\gamma = 25^\circ$ and $20^\circ$, $\mathcal{C}_{I(j_pj_n)}$ 
exhibits similar behavior. The $\mathcal{C}_{I(j_pj_n)}$ 
becomes a bit larger when $\gamma$ is reduced, because the coupling
of the rotor with the neutron hole is stronger for more axial shape. 
The discussed above structural interchange between 
the yrast and yrare bands at their (avoided) crossing is seen
between $I=13$ and 14.

For $\gamma = 15^\circ$, $\mathcal{C}_{I(j_pj_n)}$ becomes 
relatively flat with respect to spin. For $\gamma=10^\circ$  
the triaxial deformation has a minimal effect on the system. The 
concurrence is very similar to the axial case of $\gamma=0^\circ$, 
where the entanglement is large and it does not depend on $I$. 
As discussed above, the yrast and yrare bands represent 
uniform rotation about the axes perpendicular to the symmetry axis
with a different projection on the latter with the probability 
$P(\phi)=1/2\pi$ being constant. The large value of $\mathcal{C}_{I(j_pj_n)}$ 
indicates that this geometry is very different from a weak 
coupling of the rotor to the particle and hole.
 
The concurrence $\mathcal{C}_{j_p(j_nI)}$ measures the entanglement 
of the proton particle with the system rotor+neutron hole. 
For $\gamma=30^\circ$ and low $I$ the combined system represents a transverse
wobbler, which we discussed in Refs.~\cite{Q.B.Chen2022EPJA, Q.B.Chen2024PRC_v1,
Q.B.Chen2024PRC_v2}. The concurrence is only slightly above the minimal value
of 0.739 for an single fermion. There is a weak coupling caused 
by  the attractive Coriolis interactions between $\bm{J}$
and $\bm{j}_p$ as well as the repulsive recoil interaction between 
$\bm{j}_p$ and $\bm{j}_n$. The $\mathcal{C}_{j_p(Ij_n)}$ shows 
a monotonic increase with spin, which reflects the reorientation 
of the proton caused by the Coriolis interaction. For the largest
values of $I$, the entanglement is strong, which reflects 
the substantial reorientation of proton. At these values 
the neutron hole-rotor system is in the LW 
mode, which pulls the proton toward the $m$ axis as well.

For $\gamma=25^\circ$ and $20^\circ$, the triaxial potential 
keeps $\bm{j}_p$ less well aligned with the $s$ axis. 
It is more susceptible to the Coriolis interaction, which is seen 
as an increased value of $\mathcal{C}_{j_p(Ij_n)}$ at $I=8$ and the 
corresponding earlier approach of strong entanglement. The differences 
between the yrast and yrare bands reflect the differences in the 
orientation of $\bm{j}_p$ in Fig.~\ref{f:Ori}. The structural 
change at the band crossing is clearly seen as the yrast-yrare 
interchange of $\mathcal{C}_{j_p(Ij_n)}$ between $I=13$ and 14.  
Like for the $\mathcal{C}_{I(j_pj_n)}$ concurrence, the 
order of $\mathcal{C}_{j_p(Ij_n)}$ changes back to yrare 
above yrast at the upper boundary of the CR region, where 
LCV sets in.

For smaller $\gamma$, the concurrence approaches the limit of
$\gamma=0^\circ$. The $\mathcal{C}_{j_p(Ij_n)}$ starts 
large and becomes quickly spin independent. The three 
angular momenta are not restrained with respect 
to the angle $\phi$. The Coriolis interaction arranges 
them in one plane, where the angle between them decreases, 
which is reflected by the profile of $\mathcal{C}_{j_p(Ij_n)}$.
The yrast-yrare interchange of $\mathcal{C}_{j_p(Ij_n)}$ 
is not seen.

The concurrence $\mathcal{C}_{j_n(Ij_p)}$ measures 
the entanglement of the neutron hole with the 
proton particle+rotor system, which we discussed 
in Refs.~\cite{Q.B.Chen2022EPJA, Q.B.Chen2024PRC_v1, 
Q.B.Chen2024PRC_v2}. For $\gamma = 30^\circ$, 
$\mathcal{C}_{j_n(Ij_p)}$ is found to be equal to 
$\mathcal{C}_{j_p(j_nI)}$, which is expected for a symmetric 
configuration with $\mathcal{J}_s = \mathcal{J}_l$.
When triaxial deformation parameters deviates from $\gamma=30^\circ$, 
the triaxial potential still keeps $\bm{j}_n$ well aligned with 
the $l$ axis as shown in Fig.~\ref{f:Ori}. Accordingly,
the $\mathcal{C}_{j_n(Ij_p)}$ is smaller than 
$\mathcal{C}_{j_p(j_nI)}$ at the band head. 
As $I$ increases, $\mathcal{C}_{j_n(Ij_p)}$  
grows  gradually, which reflects the gradual alignment 
of $\bm{j}_n$ with the $s$-$m$ plane, where the proton+rotor system
resides. The reorientation of $\bm{j}_n$  
generated by the Coriolis interaction is counteracted 
by the potential, which binds it stronger to the $l$ axis 
when $\gamma\rightarrow 0^\circ$ (cf. Fig.~\ref{f:Ori}).
Accordingly, $\mathcal{C}_{j_n(Ij_p)}$ at $I=8$ decreases 
with $\gamma$. For $\gamma=30^\circ$, $25^\circ$, and $20^\circ$,
the yrast-yrare bands interchange their order and back with $I$. 
In contrast to $\mathcal{C}_{j_p(Ij_n)}$, the crossing spin values 
do not agree with the boundaries of the CR region, which seems 
to be related with absence of a reordering of $o_l$ there. 

 \begin{figure}[t]
  \begin{center}
    \includegraphics[width=0.90\linewidth]{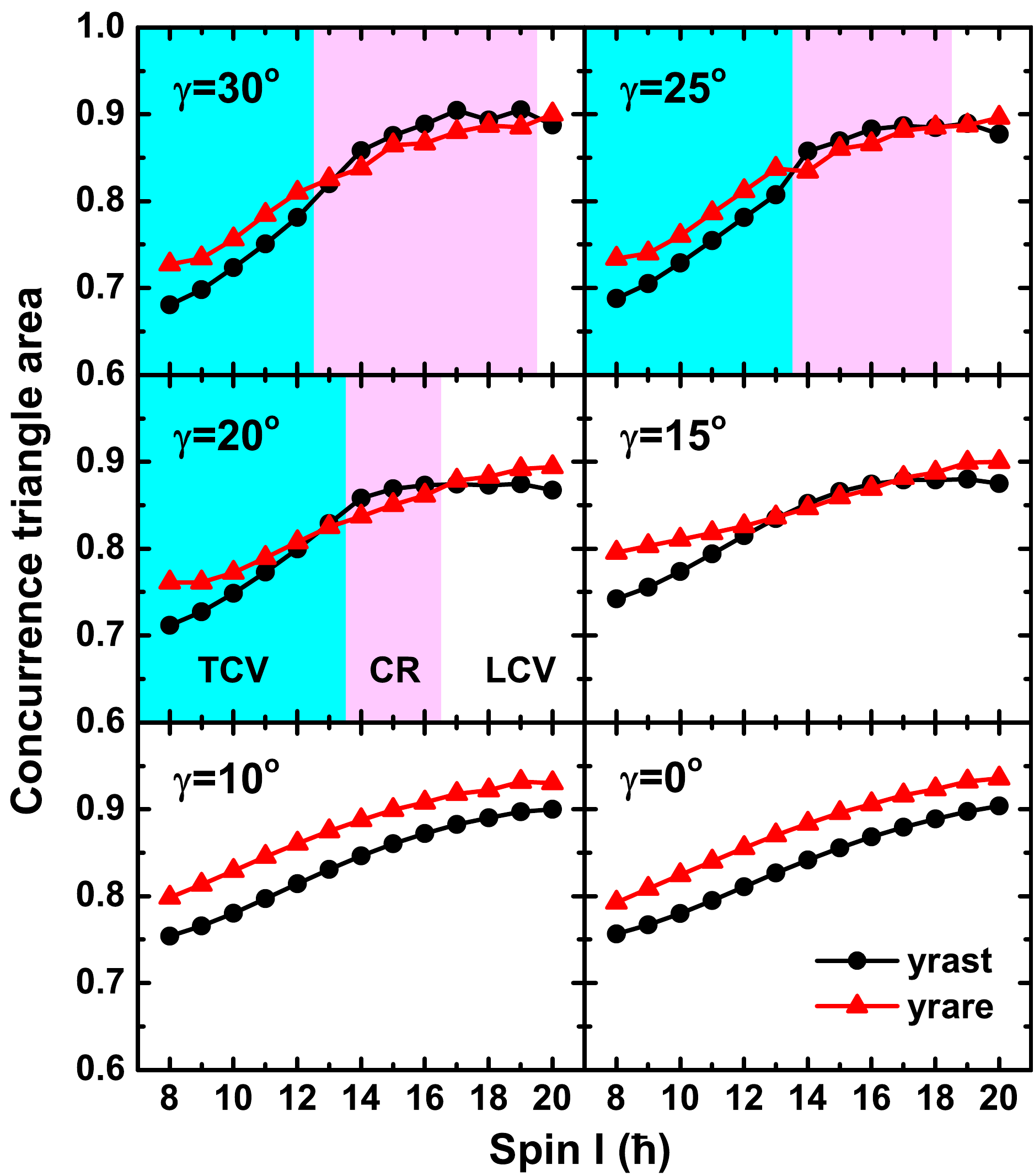}
    \caption{(Color online) The calculated area of concurrence 
    triangle as functions of spin $I$ for the yrast 
    and yrare bands with $\gamma$ ranging from $30^\circ$ to $0^\circ$ in steps 
    of $5^\circ$. The shadows denote the 
    regions of transverse chiral vibration (TCV) 
    and chiral rotation (CR), respectively.}\label{f:Area}
  \end{center}
\end{figure}

Furthermore, one notes that $\mathcal{C}_{I(j_pj_n)}$ exceeds both 
$\mathcal{C}_{j_p(j_nI)}$ and $\mathcal{C}_{j_n(Ij_p)}$. This 
can be attributed to the fact that the Coriolis interactions between 
$\bm{J}$ and $\bm{j}_p$ as well as $\bm{J}$ and $\bm{j}_n$ are both 
attractive, while the recoil interaction between 
$\bm{j}_p$ and $\bm{j}_n$ is repulsive, which leads to a
partial cancellation. In case of $\mathcal{C}_{I(j_p j_n)}$,
the recoil interaction does not contribute and combined  
Coriolis interaction with $\bm{j}_p$ and $\bm{j}_n$ causes a 
stronger entanglement.

The concurrence triangle areas $\mathcal{F}_{Ij_pj_n}$  
(\ref{eq:ConArea}) are shown in Fig.~\ref{f:Area}.
The area represents a kind of geometric average of
the lengths of its three sides, 
which we discussed in the preceding paragraphs. 
Accordingly, 
\begin{itemize}
    \item $\mathcal{F}_{Ij_pj_n}$ increases with $I$  like  
    $\mathcal{C}_{j_p(j_nI)}$ and $\mathcal{C}_{j_n(Ij_p)}$.
    \item $\mathcal{F}_{Ij_pj_n}$ increases with decreasing $\gamma$ 
    like $\mathcal{C}_{j_p(j_nI)}$.
    \item When chiral mode exists, the order between 
    yrast and yrare $\mathcal{F}_{Ij_pj_n}$ interchanges 
    at the lower border of the CR region and back to the original 
    order at upper boundary. 
    \item When chiral mode is absent, the $\mathcal{F}_{Ij_pj_n}$ 
    of the yrare remains larger than that of yrare band. 
    No yrast-yrare interchange is seen.
\end{itemize}

The triangle areas $\mathcal{F}_{Ij_pj_n}$ measures the overall
entanglement of $\bm{j}_p$,  $\bm{j}_n$ and $\bm{J}$.
From a fundamental perspective, one expects that $\mathcal{F}_{Ij_pj_n}$ increases 
with the excitation energy because, as the system excites into higher 
states, the particle and total angular momenta become more entangled, 
particularly with the onset of the CR. Additionally, 
it is anticipated that the $\mathcal{F}_{Ij_pj_n}$ will increase with spin, 
as the Coriolis interaction becomes stronger at higher angular momenta.
Figure~\ref{f:Area} confirms the expected spin-dependence.
At the band head, $\mathcal{F}_{Ij_pj_n}$ is the smallest, indicating 
the smallest entanglement. With the increase of spin $I$, $\mathcal{F}_{Ij_pj_n}$ 
increases, indicating the stronger entanglement. 

From the above analysis, one  concludes that 
the crossing behavior of concurrences triangle area can 
be considered as the fingerprint of CR. 

\subsection{Particle-particle configurations}

In this section, we examine the entanglement in the particle-particle 
configuration $\pi(1h_{11/2})^1 \otimes \nu(1h_{11/2})^1$, where both 
proton and neutron Fermi surfaces are taken as $\lambda_p = \lambda_n = e_1$.
For comparison, the entanglement in the two-proton particles configuration 
$\pi(1h_{11/2})^2$ is also investigated. The calculations use the 
same quadrupole deformation parameters, pairing gaps, and moments 
of inertia as those for the configuration $\pi(1h_{11/2})^1 \otimes 
\nu(1h_{11/2})^{-1}$. The proton and neutron Hamiltonians are
identical in the $\pi(1h_{11/2})^1 \otimes \nu(1h_{11/2})^1$ 
configuration. In this case, the proton and neutron are not 
constrained by the Pauli exclusion principle. However, the 
$\pi(1h_{11/2})^2$ configuration, where the two-proton particles, 
being identical fermions, cannot occupy the same quantum states. 
The comparisons of the two configurations demonstrate
the effects of the Pauli exclusion principle on 
the entanglement.

\begin{figure}[t]
  \begin{center}
    \includegraphics[width=0.90\linewidth]{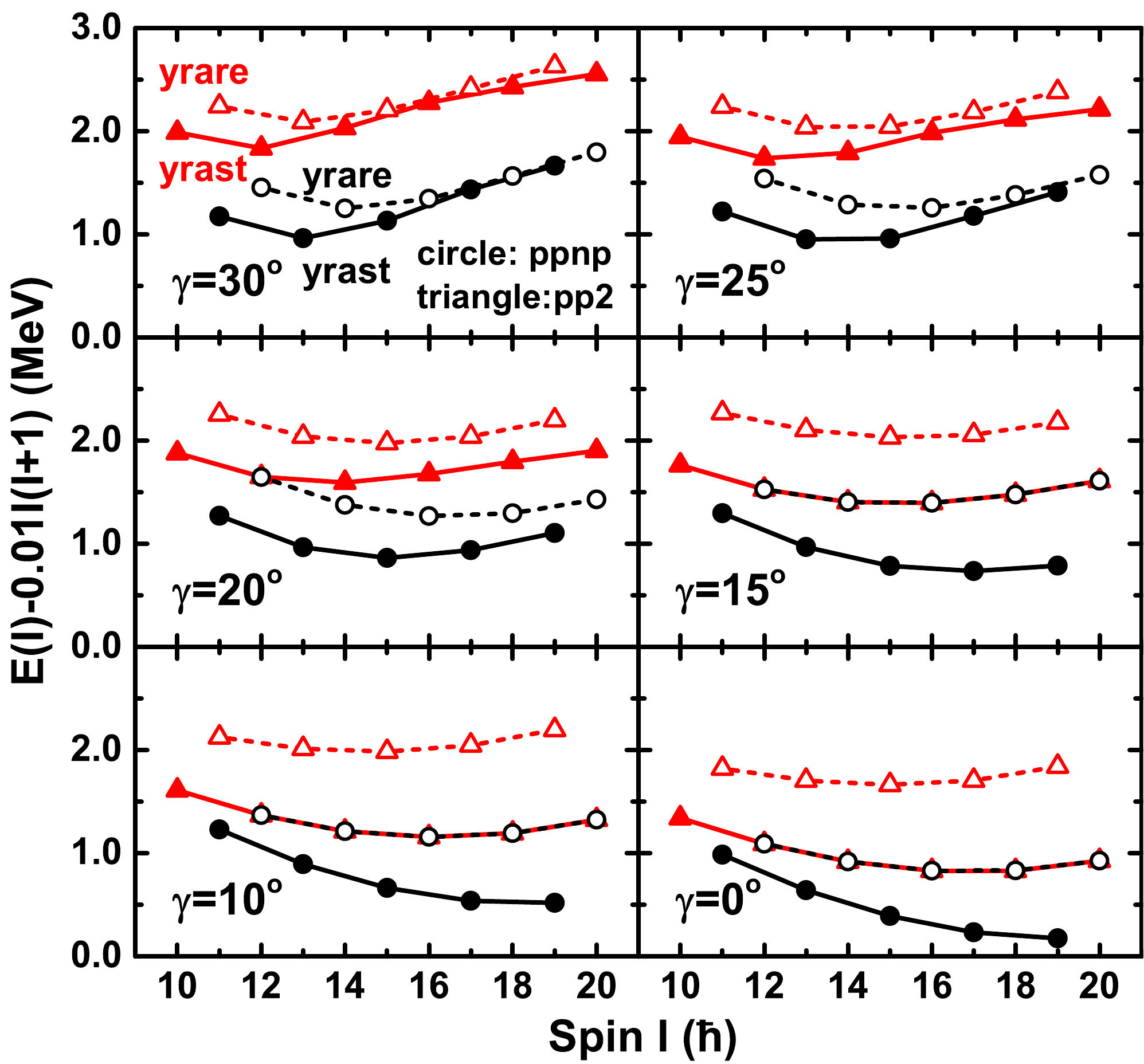}
    \caption{(Color online) PTR energies minus a common 
    rigid rotor reference as functions of spin $I$ 
    for the yrast bands of particle-particle configuration 
    $\pi (1h_{11/2})^1\otimes \nu(1h_{11/2})^1$ (labeled as ``ppnp'')
    and two-proton particles configuration $\pi (1h_{11/2})^2$ (labeled as ``pp2'') 
    with $\gamma$ ranging from $30^\circ$ to $0^\circ$ in steps 
    of $5^\circ$. }\label{f:Energy_pp_p2}
  \end{center}
\end{figure}

In Fig.~\ref{f:Energy_pp_p2}, we show the energy spectra as functions 
of spin $I$ for the yrast and yrare bands of the proton-neutron configuration 
$\pi (1h_{11/2})^1\otimes \nu(1h_{11/2})^1$ (labeled as ``pppn") 
and two-proton configuration $\pi (1h_{11/2})^2$  (labeled as ``pp2") 
with $\gamma$ ranging from $30^\circ$ to $0^\circ$ in step of $5^\circ$. 
The pp2 case was already discussed in Refs.~\cite{Frauendorf1997NPA, 
Q.B.Chen2019PRC_v1, Q.B.Chen2024PRC_v1} for large $\gamma$.
 
For both configurations, the particles are at the bottom of the shell 
and their $\bm{j}$ try to align with the $s$ axis as far as allowed
by the Pauli principle. The total angular momentum $\bm{J}$ of
rotational states is located near the $s$-$m$ plane and the states can be 
classified with respect to their signature ($\alpha=0$ for even $I$ and  
$\alpha=1$ odd $I$). In the pppn case, the yrast band is comprised of 
the odd-$I$ states because the proton and the neutron can occupy 
the most favorable quasiparticle state with the signature $\alpha=11/2$, 
which adds to 11 mod 2. In the pp2 case, the yrast band 
is comprised of even-$I$ states because the second proton 
has to occupy the quasiparticle state with 
the signature $\alpha=-11/2$ in order to obey the 
Pauli principle, which adds to 10 mod 2. 

The yrare bands have the opposite signature. For $\gamma \geq 20^\circ$, 
they represent wobbling excitations on the yrast states. Their mutual
distance initially diminishes with increasing $I$ but later increases again.
The phenomenon has been discussed in detail in Refs.~\cite{Q.B.Chen2022EPJA,
Q.B.Chen2024PRC_v1, Q.B.Chen2024PRC_v2} as
the transition from transverse wobbling (TW) at low $I$
to the flip mode (FM), and ultimately
to longitudinal wobbling (LW) at large $I$. 
At the band head, where the mean rotor energy is zero, 
the angular momentua of the $h_{11/2}$ proton and neutron 
add to $I=11\hbar$, while the angular momenta 
of the two $h_{11/2}$ protons can only provide 
$10\hbar$, due to the Pauli exclusion principle. 

For $\gamma \leq 15^\circ$, the structure rapidly approaches 
the axial limit $\gamma= 0^\circ$. The pppn odd-$I$ yrast 
band has the  proton and the neutron in the most favorable 
states with $\bm{j}$ in the plane perpendicular to 
the symmetry axis. The even-$I$ yrare band is generated
by lifting one of the nucleons to the next higher state with
$\bm{j}$ tilted somewhat out of the plane perpendicular to 
the symmetry axis. The band is antisymmetric with respect 
to exchanging protons with neutrons. The pp2 yrast band 
has the analog structure, except that both particles are 
protons. Obeying the Pauli principle, the state is antisymmetric. 
As the quasineutron and quasiproton energies and the matrix 
elements of $\bm{j}$ are the same, the states have the same energy. 

\begin{figure}[t]
  \begin{center}
    \includegraphics[width=0.90\linewidth]{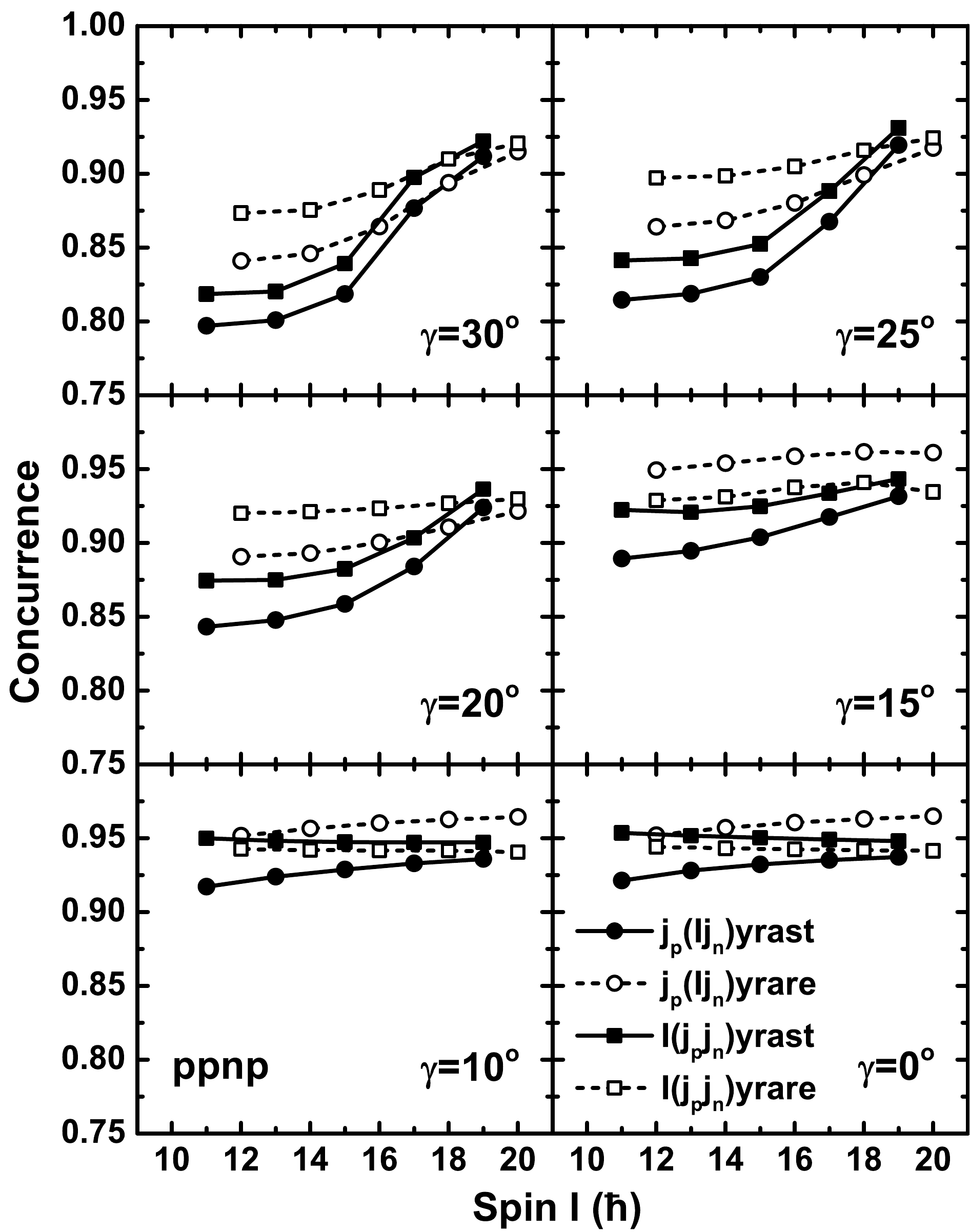}
    \caption{(Color online) The side lengths of the concurrence triangle
    for the  of particle-particle configuration 
    $\pi (1h_{11/2})^1\otimes \nu(1h_{11/2})^1$ (labeled as ``ppnp'')
    as functions of spin $I$.
    Squares display $\mathcal{C}_{I(j_pj_n)}$  and 
    circles display $\mathcal{C}_{j_n(Ij_p)}$ or $\mathcal{C}_{j_p(j_n I)}$.  
    The full line with solid symbol and dashed line 
    with empty symbol show the results for the yrast and yrare bands,
    respectively.
    }\label{f:Concurrence_ppnp}
  \end{center}
\end{figure}

One can see the identical energies from a complementary perspective.
The considered configuration of $\pi (1h_{11/2})^1 
\otimes \nu (1h_{11/2})^1$ is expected 
on the $Z=N$ line. The odd-$I$ yrast configuration has isospin 
$T=0$. The even-$I$ yrare band has $T=1$, $T_3=0$. 
The even-$I$ yrast band of $\pi (1h_{11/2})^2$ has 
$T=1$, $T_3=-1$. Being the isobar analog state 
it has the same energy. The analog case of the  $\pi (1g_{9/2})^1 
\otimes \nu (1g_{9/2})^1$ configuration in $^{70}$Br has 
been studied in Ref.~\cite{Jenkins2002PRC} 
using the two-quasiparticle + axial rotor model.

\begin{figure}[t]
  \begin{center}
    \includegraphics[width=0.90\linewidth]{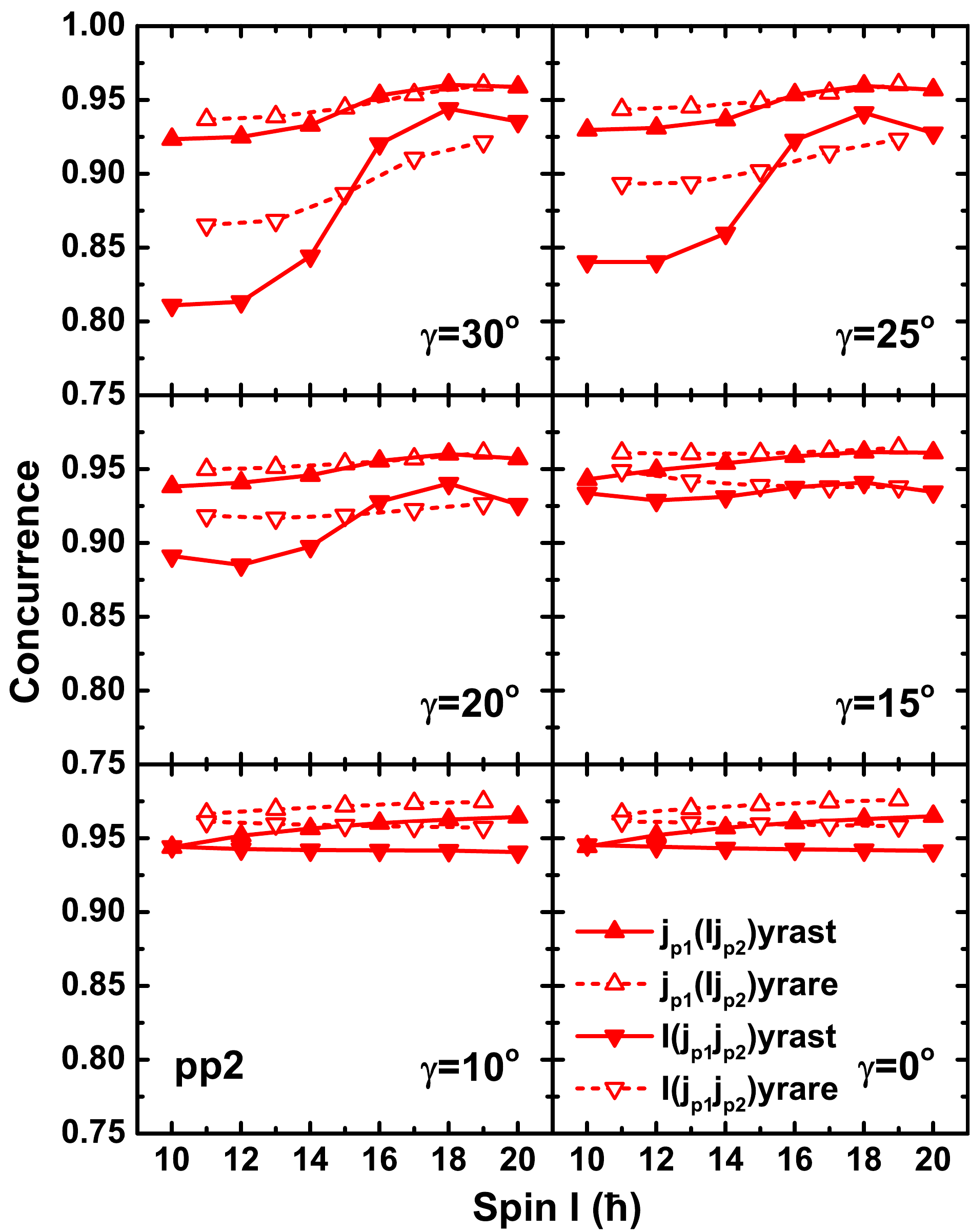}
    \caption{(Color online) The side lengths of the concurrence triangle
    for the  of particle-particle configuration  $\pi 
    (1h_{11/2})^2$ (labeled as ``pp2'') as functions of spin $I$.
    Triangle down display 
    $\mathcal{C}_{I(j_{p1}j_{p2})}$ and triangle up
    displays $\mathcal{C}_{j_{p1}(j_{p2} I)}$ or $\mathcal{C}_{j_{p2}(j_{p1} I)}$. 
    The full line with solid symbol and dashed line 
    with empty symbol show the results for the yrast and yrare bands,
    respectively.
    }\label{f:Concurrence_pp2}
  \end{center}
\end{figure}

In Figs.~\ref{f:Concurrence_ppnp} and \ref{f:Concurrence_pp2}, 
we present the side lengths of the concurrence triangle in the 
ppnp and pp2 configurations, respectively. For ppnp configuration, 
the proton and neutron Hamiltonian are identical, hence 
the lengths of the concurrence triangles $\mathcal{C}_{j_p(j_n I)}$ 
is identical with $\mathcal{C}_{j_{n}(Ij_{p})}$. 
Moreover, for pp2 configuration, the two protons 
are indistinguishable. If we label one of them as $p1$ and
the other one as $p2$, $\mathcal{C}_{j_{p1}(j_{p2} I)}$
is identical with $\mathcal{C}_{j_{p2}(Ij_{p1})}$. Hence,
in Figs.~\ref{f:Concurrence_ppnp} and \ref{f:Concurrence_pp2}, 
we only show the results of $\mathcal{C}_{j_p(j_n I)}$ and 
$\mathcal{C}_{j_{p1}(j_{p2} I)}$, respectively. 

\begin{figure}[ht]
  \begin{center}
    \includegraphics[width=0.90\linewidth]{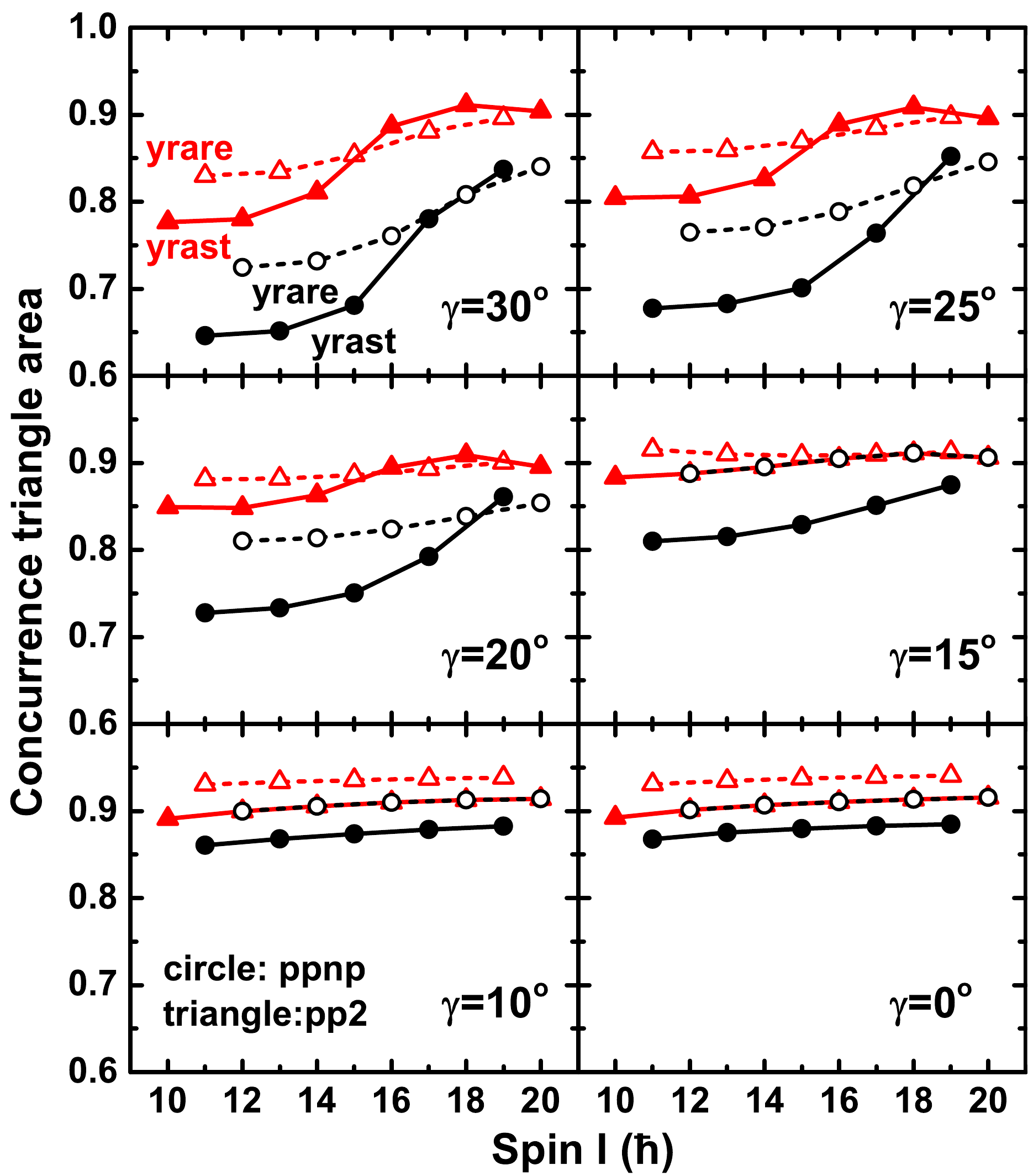}
    \caption{(Color online) The calculated area of concurrence triangle,
    as functions of spin $I$ for the yrast bands of particle-particle configuration 
    $\pi (1h_{11/2})^1\otimes \nu(1h_{11/2})^1$ (labeled as ``ppnp'') 
    and two-proton particles configuration $\pi (1h_{11/2})^2$ 
    (labeled as ``pp2'') with $\gamma$ ranging from $30^\circ$ to $0^\circ$ in steps 
    of $5^\circ$. }\label{f:Area_pp_p2}
  \end{center}
\end{figure}

For the pppn configuration, the concurrence $\mathcal{C}_{I(j_pj_n)}$ 
measures the entanglement of the rotor with the proton-neutron pair. 
For $\gamma=30^\circ$, the pair is not strongly entangled at low $I$, 
in contrast to chiral proton-neutron hole system,
which is strongly entangled with the rotor right from $I=8$. 
The reason for the difference is not obvious. In both cases a substantial 
reorientation of the  rotor to respectively the $s$ axis the tilted axis
at $45^\circ$ in the $s$-$l$ is involved.  This means that in 
the case of the proton-neutron pair the resulting state is close 
to a product of a rotor and a two-particle state, while for 
the proton particle-neutron hole case such an approximation 
does not hold. Naturally, the entanglement 
of the proton-neutron pair grows with $I$.

The concurrence $\mathcal{C}_{j_p(Ij_n)}$ measures the entanglement of 
the proton with the rotor-neutron system. In the case of 
$\gamma=30^\circ$, the yrast band starts at a low
value. The proton acts just as a spectator to the odd-neutron transverse
wobbler. At larger $I$, the TW mode couples by the Coriolis interaction
to the proton and the entanglement grows. Following the general trend,
the yrare band starts with a somewhat larger $\mathcal{C}_{j_p(Ij_n)}$,
which increases with $I$. 

The concurrence $\mathcal{C}_{I(j_{p1}j_{p2})}$ measures 
the entanglement of the rotor with the proton pair in the 
pp2 configuration. For $\gamma=30^\circ$, 
similar to the proton-neutron pair, 
the yrast band of the proton pair starts with low entanglement which increases 
with $I$. The yrare band starts at higher concurrence to grow with $I$. 
The two-proton case has been extensively investigated in 
Ref.~\cite{Q.B.Chen2024PRC_v2}, where the purity
figures represent the concurrence in essence.

The concurrence $\mathcal{C}_{j_{p1}(Ij_{p2})}$ measures 
the entanglement of one of proton with the other 
proton and rotor system. For $\gamma=30^\circ$,
the concurrence $\mathcal{C}_{j_{p1}(Ij_{p2})}$ for the 
yrast starts with the large value of 0.925, 
which leaves little margin to for further growth with $I$. 
The concurrence of the yrare band is quite similar. 
These values are substantially larger then the concurrences 
$\mathcal{C}_{j_{p}(Ij_{n})}$ of the proton-neutron pair. 
The difference reflects the additional entanglement 
caused by the Pauli Principle. 

For $\gamma=25^\circ$ and $20^\circ$, the concurrence patterns 
in the pppn and pp2 configurations are similar to $\gamma=30^\circ$, 
where the TW features getting washed out and 
the entanglement increases with decreasing triaxiality. For 
$\gamma \leq 15^\circ$, the axial limit $\gamma=0^\circ$ is approached.
The angular momenta  $\bm{j}_p$, $\bm{j}_n$,  and $\bm{J}$ are
not constraint with respect to the azimuthal angle $\phi$. The Coriolis 
interaction aligns $\bm{j}_p$ and $\bm{j}_n$ with $\bm{J}$, which 
is reflected by the large concurrences. As $\gamma$ decreases, 
the concurrences in the two configurations increase.
The yrare concurrences of pppn agree with 
the yrast concurrences of pp2, because 
the states have isobar analog structures, 
which we discussed above.

The concurrence triangle areas (\ref{eq:ConArea}) 
$\mathcal{F}_{Ij_pj_n}$ for ppnp configuration 
and $\mathcal{F}_{Ij_{p1}j_{p2}}$ for pp2 configuration
are shown in Fig.~\ref{f:Area_pp_p2}.
The area represents a kind of geometric average of
the lengths of its three sides, 
which we discussed in the preceding paragraphs. The figure demonstrates
most clearly that the anti-symmetrization of the two-proton states
systematically increases the entanglement between the three angular momenta. 
Additionally, as $\gamma$ decreases, 
the concurrence triangle area associated with the two 
configurations increases. This suggests a stronger 
entanglement between the particles and core states
when $\gamma$ becomes smaller, as shown in 
Figs.~\ref{f:Concurrence_ppnp} and \ref{f:Concurrence_pp2}.

\section{Summary}

In summary, we have investigated the entanglement between the 
total angular momentum and two-quasiparticle angular momenta,
using the PTR model. As study cases, we examine the configurations 
$\pi (1h_{11/2})^1 \otimes \nu (1h_{11/2})^{-1}$ which involves
a single proton particle and a single neutron hole, 
$\pi (1h_{11/2})^1 \otimes \nu (1h_{11/2})^{1}$ which involves
a single proton particle and a single neutron particle, 
as well as $\pi (1h_{11/2})^2$ which involves
two-proton particles, coupled to a triaxial rotor. 
The analyses were conducted by partitioning the 
coupled system into three subsystems, each described by 
its corresponding reduced density matrix.

The appearance of a chiral mode in the particle-hole configuration 
$\pi (1h_{11/2})^1 \otimes \nu (1h_{11/2})^{-1}$ 
and of the transverse wobbling mode in particle-particle mode 
$\pi (1h_{11/2})^2$ have been demonstrated by 
the probability distributions via SCS maps and 
SSS plots in our previous publications~\cite{Frauendorf2015Conf, 
Q.B.Chen2022EPJA, Q.B.Chen2024PRC_v1, Q.B.Chen2024PRC_v2}. 
Additional SCS maps and SSS plots have been added to study the
disappearance of the modes with decreasing triaxiality. The
information on the angular momentum geometry was complemented by
plots of the orientation parameters, which are the expectation values 
of the squared angles of the angular momenta with respect to the three
principal axis of the triaxial shape.

In the case of particle-hole configuration, our results 
demonstrate that for $\gamma = 30^\circ$, $25^\circ$, and $20^\circ$,
the rotational mode transitions from transverse chiral vibration (TCV) with 
respect to the $s$-$l$ plane to chiral rotation (CR), and ultimately
to a longitudinal chiral vibration (LCV) with respect to the $s$-$m$ plane.  
For cases with small triaxial deformations
 $10^\circ$ and $0^\circ$, the collective chiral modes 
 are absent. The lowest bands represent 
 different rotating quasiparticle configurations. 
 The case of $\gamma = 15^\circ$ has transitional 
 character between the two regimes.
 
In the case of the particle-particle configuration,
our results demonstrate that for $\gamma = 30^\circ$, 
$25^\circ$, and $20^\circ$, the rotational 
mode transitions from transverse wobbling (TW) with 
respect to the $s$ axis to the flip mode (FM), and ultimately
to a longitudinal wobbling (LW) with respect to the $m$ axis.  
For cases with small triaxial deformations, 
$10^\circ$, and $0^\circ$, the collective wobbling modes 
are absent. The lowest bands represent different
rotating quasiparticle configurations. 
The case of $\gamma = 15^\circ$ has transitional 
character between the two regimes.
 
Entanglement was quantified by the concurrence triangle. Its sides 
have a length given by the three bipartite concurrences 
$\mathcal{C}_{I(j_p j_n)}$, $\mathcal{C}_{j_p(j_n I)}$, and 
$\mathcal{C}_{j_n(I j_p)}$, which, respectively, quantify
the entanglement between the total angular moment $\bm{J}$ 
and the subsystem $(\bm{j}_p\bm{j}_n)$ composed of the proton 
angular momentum $\bm{j}_p$ and the neutron particle 
or hole angular momentum $\bm{j}_n$, the entanglement between 
$\bm{j}_p$ and the subsystem $(\bm{j}_n\bm{J})$, 
and the entanglement between $\bm{j}_n$ and 
the subsystem $(\bm{J}\bm{j}_p)$. The overall entanglement 
between the three angular momenta is measured by 
the area $\mathcal{F}_{Ij_pj_n}$ of the triangle, 
which represents a geometric average of the 
tree bipartite concurrences. In the case of 
the two-proton configuration, the bipartition 
was carried out among the two-proton 
angular momenta $\bm{j}_{p1}$ and $\bm{j}_{p2}$ 
and the total angular momentum $\bm{J}$.

In the chiral regime ($\gamma\geq 20^\circ$) of the $\pi (1h_{11/2})^1 
\otimes \nu (1h_{11/2})^{-1}$ configuration, the concurrence 
area $\mathcal{F}_{Ij_pj_n}$ rises with spin $I$, reflecting 
increasing entanglement caused by the Coriolis coupling between 
$\bm{j}_p$, $\bm{j}_n$, and $\bm{J}$. It has a maximum 
in the CR region, where the three sides of the concurrence triangles, 
$\mathcal{C}_{I(j_p j_n)}$, $\mathcal{C}_{j_p(j_n I)}$, and 
$\mathcal{C}_{j_n(I j_p)}$ have about the same lengths.
At low $I$ in the CV region, $\mathcal{F}_{Ij_pj_n}$ of the
yrast band is smaller than the one of the yrare one. 
When the system enters the CR region the order reverses. 
At end of the CR region the system returns to the original 
order. The two crossings of $\mathcal{F}_{Ij_pj_n}$ as function 
$I$ are caused by corresponding order changes of the bipartite 
concurrence $\mathcal{C}_{j_p(j_n I)}$, which can be traced 
back to a reorientation of $\bm{j}_p$. 

In the considered examples the chiral symmetry is moderately broken,
which leaves coupling terms between the left- and right-handed configurations.
Restoring the chiral symmetry by forming even and odd linear combination results in 
somewhat different orientations of $\bm{j}_p$. The CR appears as the crossing
between the CV yrare with the CV yrast bands, which keep their structure 
(the angle $\phi_p$ of $\bm{j}_p$ with the $s$-$l$ plane) through the crossing. 
This leads to an interchange of $\phi_p$ (and other properties as well). 
Above the upper boundary of the CR region, $\phi_p$ becomes larger 
than $45^\circ$, which interchanges the structural order again. 
Therefore, the crossing behavior of the concurrence triangle area, 
but other features as the energies and electromagnetic transition 
matrix elements as well, can be considered a characteristic 
signature of CR in the cases of moderate breaking of 
the chiral symmetry, which have been identified so far.
The two crossing bands differ from each other not only by 
the phase between the left- and right- handed configurations, 
as in the case of strong chiral symmetry breaking, but also 
by the orientation of $\bm{j}_p$.

In the wobbling regime ($\gamma\geq 20^\circ$) of the particle-particle 
configuration $\pi(1h_{11/2})^1 \otimes \nu(1h_{11/2})^1$ 
and the two-proton particles configuration $\pi(1h_{11/2})^2$, 
the concurrence area $\mathcal{F}_{Ij_pj_n}$
rises with spin $I$, reflecting increasing entanglement caused 
by the Coriolis coupling between $\bm{j}_p$, $\bm{j}_n$, and
$\bm{J}$. The area of the two-proton configuration is always larger than the
area of the proton-neutron configuration. The extra 
entanglement is caused by the Pauli exclusion principle between 
the two protons, because the bipartite concurrence 
$\mathcal{C}_{j_{p1}(Ij_{p2})}$ of the two protons is 
large for all $I$, while the bipartite concurrence 
$\mathcal{C}_{j_{p}(Ij_{n})}$ of the proton-neutron 
pair starts with a low value to increase with $I$.

In the near-axial regime ($\gamma<15^\circ$), 
the collective chiral or wobbling modes are absent 
in the lowest bands, which represent configurations of rotating
quasiparticles. All the concurrences are large and depend 
weakly on $I$. The strong entanglement between 
the angular momenta $\bm{j}_{p}$, $\bm{j}_{n}$, 
and $\bm{J}$ arises because the axial potential 
cannot restrain their angle $\phi$ with the $s$-$l$ plane
and the Coriolis interaction locates them in
a plane containing the $l$ axis.

The conclusions drawn in this study are, to some extent, 
specific to the PTR model, which incorporates only the Coriolis 
and recoil interactions. Despite of this, many of the features 
 are a consequence of the relatively small dimensionality 
of the entangled Hilbert spaces and are expected to 
apply to other coupled systems with similar dimensions.
It seems interesting to extend the present study to more 
complicated chiral modes with three-quasiparticles
or four-quasiparticles, which will require higher-dimensional 
concurrence measures.

\section*{Acknowledgments}

One of the authors (Q.B.C.) thanks Dong Bai for
helpful discussions on describing the entanglement 
of three body system. This work was supported by the National 
Key R\&D Program of China No.~2024YFE0109803 and the National 
Natural Science Foundation of China under Grant No.~12205103.


%

\end{document}